\documentclass[prl,twocolumn,english,superscriptaddress,nobalancelastpage]{revtex4-2}
\pdfoutput=1 % keep for arxiv, remove for physical review submission

\let\SJtextoslash\o % keep the text "ø" (stylesheet redefines \o as \omega); restored before the bibliography
\usepackage{stylesheet}
\usepackage{graphicx,subcaption}
\usepackage[font=small,labelfont=bf,justification=Justified]{caption}
\usepackage{dsfont}
\usepackage{tikz}
\usetikzlibrary{positioning,shadings,decorations.pathmorphing}

\newcounter{appsection}
\renewcommand{\theappsection}{S\arabic{appsection}}
\newcommand{\appsection}[1]{%
  \refstepcounter{appsection}%
  \section*{\theappsection. #1}%
  \setcounter{theorem}{0}%
  \setcounter{figure}{0}%
  \setcounter{table}{0}%
  \setcounter{equation}{0}%
  \renewcommand{\thetheorem}{\theappsection.\arabic{theorem}}%
  \renewcommand{\thefigure}{\theappsection.\arabic{figure}}%
  \renewcommand{\thetable}{\theappsection.\arabic{table}}%
  \renewcommand{\theequation}{\theappsection.\arabic{equation}}%
  \renewcommand{\theHtheorem}{app.\theappsection.\arabic{theorem}}%
  \renewcommand{\theHfigure}{app.\theappsection.\arabic{figure}}%
  \renewcommand{\theHtable}{app.\theappsection.\arabic{table}}%
  \renewcommand{\theHequation}{app.\theappsection.\arabic{equation}}%
}

\begin{document}

\notoc

%% ======================================================================
%% Title, authors, abstract
%% ======================================================================

\title{Monolithic Quantum Error Correction in the Presence of Distinguishability}

\author{Shubham~P.~Jain}
\email{shubhamj810@gmail.com}
\affiliation{\QUICS}

\author{Eric~R.~Hudson}
\affiliation{\CQSE}
\affiliation{\UCLA}

\author{Wesley~C.~Campbell}
\affiliation{\CQSE}
\affiliation{\UCLA}

\author{Alexey~V.~Gorshkov}
\affiliation{\QUICS}
\affiliation{\JQI}

\author{Victor~V.~Albert}
\affiliation{\QUICS}

\date{\today}

\begin{abstract}
 Quantum error-correcting codes require that the environment cannot distinguish relevant error processes acting on different codewords. In practice, residual Zeeman, Stark, or anharmonic interactions shift the underlying states, making errors distinguishable and weakening the conditions for perfect recovery. We study this effect with an analytically solvable four-level model and a canonical spontaneous-emission noise model for atoms and molecules. Distinguishability raises the effective Kraus rank of the noise channel, lowers the recovered fidelity, and sets an operational checking time that shortens as the frequency separation between the emitted photons increases. We find that rapid syndrome checking suppresses the buildup of distinguishability and quantify how fast one needs to check to restore near-perfect recovery. Error distinguishability is thus a practical limitation of monolithic error correction, and our analysis identifies fast syndrome extraction as a route to mitigating it.
\end{abstract}

\maketitle

%% ======================================================================
%% Introduction
%% ======================================================================
Fault-tolerant overhead grows rapidly with the physical error rate~\cite{fowler2012surface,aharonov1997faulttolerant,dennis2002topological,acharya2023suppressing}, so quantum error correction (QEC) architectures need to be built from the best physical qubits possible. One route to lower physical error rates is monolithic encoding, which stores information in the multilevel structure of a single system, such as a bosonic mode~\cite{gottesman2001encoding,michael2016newa,leghtas2013hardwareefficienta}, molecule~\cite{albert2020robust,jain2024absorptionemission}, or atom~\cite{albert2020robust,jain2024absorptionemission,aydin2025class,gross2021designing,omanakuttan2023multispin,kubischta2023family}, so that the dominant noise processes are correctable and sub-dominant errors set the residual error floor. 
The encoded qubits then serve as the physical qubits of a larger fault-tolerant architecture with a lower effective error rate~\cite{noh2020faulttoleranta,putterman2025hardwareefficient}.

A common ingredient in such constructions is the indistinguishability of code states to the environment, so that an error reveals only that a decay occurred, not which physical basis state within the encoded superposition underwent the decay.
We call error processes that the environment cannot tell apart in this way \emph{degenerate}.
The transition frequencies of such processes coincide, which can happen when the underlying levels are degenerate, as in atomic and molecular manifolds, or uniformly spaced, as in a harmonic oscillator.
In real devices, Zeeman or Stark shifts and anharmonicities such as the Kerr nonlinearity perturb these transition frequencies, imprinting which-path information on the environment and degrading recoverability. This is the error-correction counterpart of the complementarity tradeoff between which-path information and interference visibility~\cite{scully1991quantum,englert1996fringe}. We ask how error-correction performance changes as an initially correctable error process becomes progressively distinguishable to the environment.

Operationally, an error channel is described by Kraus operators $\{E_k\}$, which are correctable if all states $|\psi\ket,|\phi\ket$ in the codespace $\mathcal{C}$ satisfy the Knill--Laflamme (KL) conditions~\cite{knill1997theorya}
\begin{equation}
 \bra\psi|E_a^\dagger E_b^{}|\phi\ket=c_{ab}\bra\psi|\phi\ket,
\end{equation}
where $c_{ab}$ are the elements of a Hermitian matrix. The state of the environment after the noise is set by the overlaps $\bra\psi|E_a^\dagger E_b^{}|\psi\ket$. Since these equal the same $c_{ab}$ for every code state, the environment learns which error occurred but nothing about the encoded state. In the fully distinguishable limit, a resolved transition $E=|f\ket\bra s|$ from a source state $|s\ket$ to an orthogonal state $|f\ket$ rules out exact correction on any codespace with support on $|s\ket$~\cite{jain2024absorptionemission}, while partial resolution leads to the approximate degradation studied here. The same issue arises for bosonic codes under Kerr anharmonicity~\cite{albert2018performancea,cai2021bosonica,brady2024advancesa}, and for decoherence-free subspaces and noiseless subsystems~\cite{zanardi1997noiseless,lidar1998decoherencefree,knill2000theoryb}, whose passive protection degrades once the symmetry of the system--bath coupling is broken~\cite{bacon1999robustness,lidar2000protecting}.

Distinguishable errors add effective Kraus operators and break the KL conditions. This is intuitively harmful, but a quantitative, operational characterization is still needed, especially for small energy shifts that only partially lift the degeneracy of the error processes.
Once the KL conditions cease to hold exactly, the problem naturally falls within the existing framework of approximate quantum error correction, where useful protection can persist and near-optimal recovery maps still provide meaningful performance guarantees~\cite{leung1997approximate,schumacher2001approximate,beny2010general,ng2010simple}.

Here we show that a small energy shift $\Delta$ does not spoil error correction at once: which-path information leaks to the environment gradually, on the timescale $1/\Delta$, and the loss of recoverable fidelity grows with the shift. This sets a natural clock, since correcting errors faster than the leakage restores near-perfect protection.
We quantify this effect through the number of distinguishable error components and the checking time required to maintain a target fidelity.
As a second contribution, we develop a canonical noise model for atomic and molecular angular-momentum systems. We present an electric-dipole spontaneous-emission channel whose jump operators are fixed by angular-momentum selection rules, with platform-specific reduced matrix elements absorbed into a small number of effective parameters.
Multilevel spontaneous-emission master equations are standard in quantum optics~\cite{scully2012quantum,dum1992monte,molmer1993monte}. Our construction turns them into a compact, platform-independent QEC error model that covers the full angular-momentum ladder and partially resolved transitions.

\begin{figure}[t]
  \begin{subfigure}[t]{\columnwidth}
    \caption{}
    \label{fig:4lvl}
    \centering
    \includegraphics[width=0.56\columnwidth]{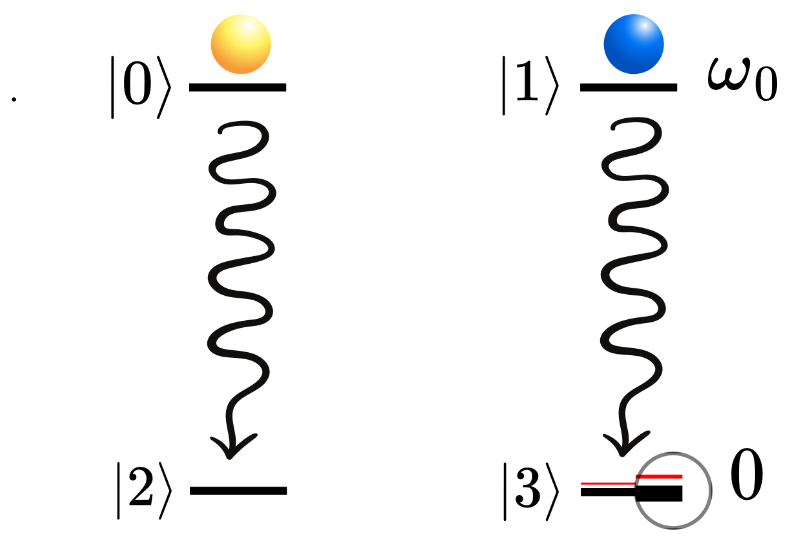}
  \end{subfigure}\par\medskip

  \begin{subfigure}[t]{\columnwidth}
    \caption{}
    \label{fig:rotor}
    \includegraphics[width=\columnwidth]{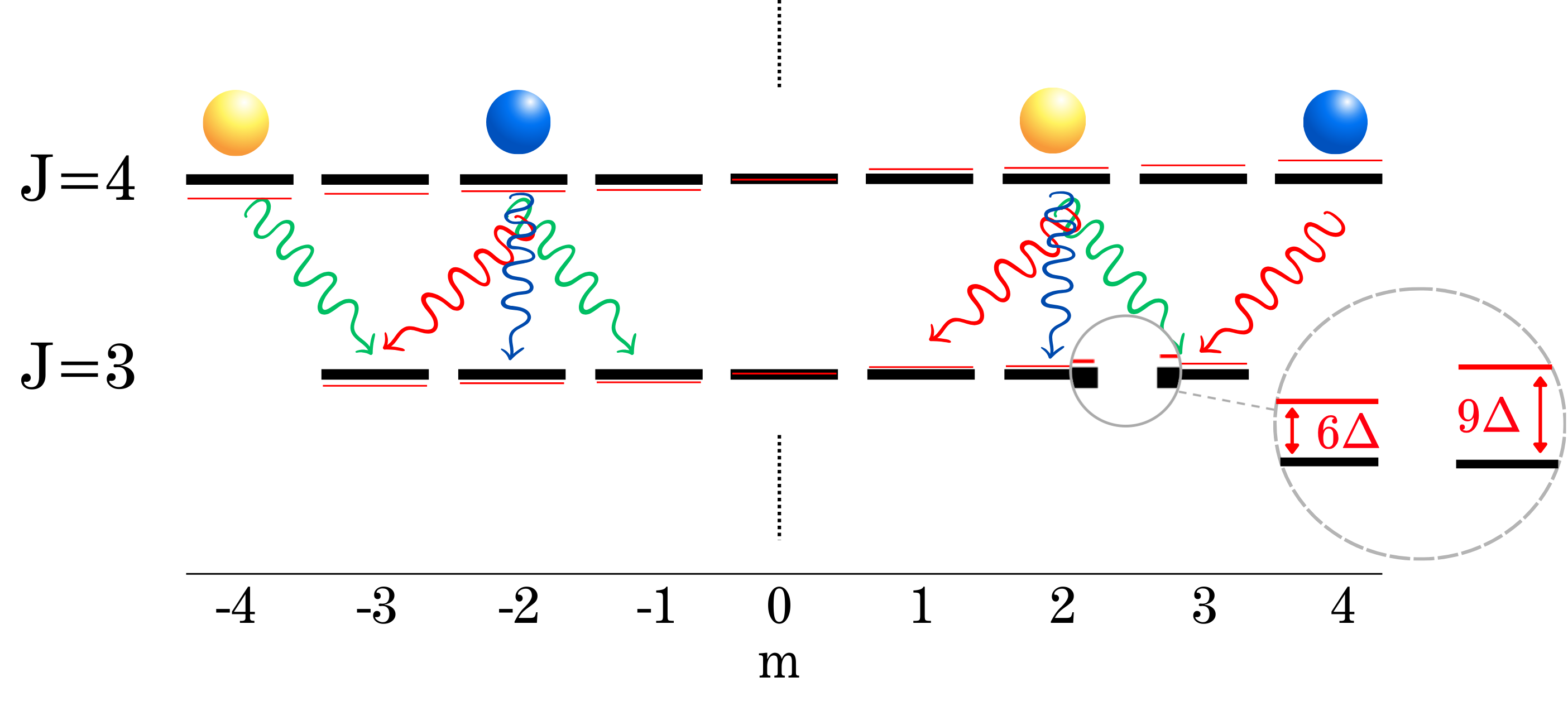}
  \end{subfigure}

  \caption{(a) The four-level model with the degeneracy-lifting term $\Delta|3\ket\bra3|$ and (b) a representative atomic system with the degeneracy-lifting term $\sum_{J,m}\Delta\,J\,m\,|J,m\ket\bra J,m|$. Wavy arrows show the jump operators of the corresponding noise models. Yellow and blue spheres mark the basis states of the codes hosted in each system. Black lines show the ideal degenerate spectrum and red lines the lifted one.
  }
  \label{fig:systems}
\end{figure}

%% ======================================================================
%% Four-level system
%% ======================================================================
\paragraph{Four-level system.} We begin by studying a four-level system with the Hamiltonian
\begin{equation}\label{eqn:4-level-hamiltonian}
    H_\mathcal{S} = \omega_0\left (|0\ket \bra0| + |1\ket\bra 1|\right ),
\end{equation}
shown in \cref{fig:systems}(a).

Decay is described by a Lindbladian with the jump operator $\sqrt{\gamma}\left(|2\ket\bra0|+|3\ket\bra1|\right)$, where $\gamma$ is the decay rate. We use units with $\hbar=1$ throughout. The codespace is the span of $\{|0\ket,|1\ket\}$: an encoded state either stays there or decays coherently to the span of $\{|2\ket,|3\ket\}$. The natural recovery $\mathcal{R}_4$ measures whether a decay occurred, does nothing on the null outcome, and maps $|2\ket\mapsto|0\ket$ and $|3\ket\mapsto|1\ket$ on the decay outcome. This picture fails once the $|0\ket\rightarrow|2\ket$ and $|1\ket\rightarrow|3\ket$ transitions are made distinguishable. We quantify this with the \emph{degeneracy-lifting strength} (or simply \emph{lifting strength}) $\Delta$ through an additional term in the Hamiltonian (\cref{fig:systems})
\begin{equation}
    H_\Delta=\Delta\, |3\ket\bra 3|.
\end{equation}

The recovery $\mathcal{R}_4$ gives a direct state-level view of the mechanism. 
For an initial code state $|\psi\ket$, let $\rho_{\mathrm{d}}(t)$ be the normalized post-recovery state conditioned on a decay outcome, i.e., the decayed branch after $\mathcal{R}_4$ has mapped it back to the codespace.
The branch-averaged decoded state fidelity is
\begin{equation}\label{eqn:state-fidelity-4level-system}
F_\psi(t)=p_{\mathrm{null}}+p_{\mathrm{decay}}\bra\psi|\rho_{\mathrm{d}}(t)|\psi\ket ,
\end{equation}
where $p_{\mathrm{null}}=e^{-\gamma t}$ and $p_{\mathrm{decay}}=1-e^{-\gamma t}$. 
The null branch is harmless after normalization. All loss comes from the decay branch, where the degeneracy-lifting term lets relative phase accumulate between the two decay paths. At short times, the decay probability contributes one power of $\gamma\, t$, while this phase mismatch contributes two powers of $\Delta\, t$. Thus, for every code state,
\begin{equation}
1-F_\psi(t)=c_\psi\gamma\Delta^2t^3+\mathcal{O}(t^4),
\end{equation}
where, as derived in Supplemental Material~\ref{sm:4lvl}, the state-dependent coefficient is $c_\psi=|a|^2|b|^2/3 \leq 1/12$ for $|\psi\ket=a|0\ket+b|1\ket$.
This expansion yields the two perfectly recoverable limits: if $\gamma=0$, no decay occurs, while if $\Delta=0$, a decay carries no which-path information.
To reveal the mechanism behind this effect and make the analysis state-independent, we next introduce two diagnostic tools: the \emph{checking time} associated with a given encoding and the \emph{Kraus rank} of the noise channel.

%% ======================================================================
%% Diagnostics
%% ======================================================================
For a target fidelity $F_0\in(0,1]$, we define the checking time of a code, noise channel, and recovery as $\tau(F_0)=\inf\{t\,|\,F_\mathcal{R}(t)<F_0\}$, where $F_\mathcal{R}(t)$ is the entanglement fidelity of the recovered channel $\mathcal{R}\circ\mathcal{N}_t$ on the code and $\mathcal{N}_t$ is the noise accumulated over time $t$. It is the earliest time at which the recovered fidelity drops below $F_0$, and hence sets how often error correction must be performed to preserve the encoded information for the specified recovery. We take $\mathcal{R}$ to be the transpose, or Petz, recovery $\mathcal{R}_{\mathrm{tr}}$~\cite{petz1986sufficient,petz1988sufficiency,barnum2002reversing,ng2010simple}, for two reasons. State-level fidelities such as $F_\psi$ are available in closed form only for solvable models, whereas the transpose map is defined for any code and channel. Its fidelity also bounds that of the optimal recovery, $1-F_{\mathrm{opt}}\le 1-F_{\mathrm{tr}}\le 2(1-F_{\mathrm{opt}})$~\cite{barnum2002reversing,fletcher2007optimuma,ng2010simple,tyson2010twosided,mandayam2012unified,zheng2024nearoptimal}, so the checking time reflects the code rather than the decoder.

For Kraus operators $\{E_l\}$ and code basis states $\{|\bar\nu\ket\}$, the transpose fidelity is~\cite{zheng2024nearoptimal} 
\begin{equation}\label{eqn:transpose-fidelity}
    F_{\mathrm{tr}} = \frac{1}{d_{\mathrm{L}}^2}\lVert\Tr_{\mathrm{L}} \sqrt{M}\rVert_{\mathrm{F}}^2,
\end{equation}
where $d_{\mathrm{L}}$ is the logical dimension, $M_{[\mu,l],[\nu,k]}=\bra\bar\mu|E_l^\dagger E_k^{}|\bar\nu\ket$ is the QEC matrix, $\Tr_{\mathrm{L}}$ traces over the logical index, and $\lVert\cdot\rVert_{\mathrm{F}}$ is the Frobenius norm. It is the entanglement fidelity of $\mathcal{R}_{\mathrm{tr}}\circ\mathcal{N}$ on the code and determines the state fidelity averaged over Haar-random logical states, $\bar F=(d_{\mathrm{L}}F_{\mathrm{tr}}+1)/(d_{\mathrm{L}}+1)$~\cite{horodecki1999general,nielsen2002simple}.

The Kraus rank of a channel is the minimal number of Kraus operators in any representation, equivalently the rank of its Choi matrix~\cite{watrous2018theory}. We use it as a channel-side count of the independent error components the environment can resolve, which grows as degenerate decay processes split into distinguishable ones.

In the four-level model, any $\Delta\neq0$ raises the Kraus rank from $2$ to $3$ for all $t>0$ (End Matter): the decay paths $|0\ket\rightarrow|2\ket$ and $|1\ket\rightarrow|3\ket$ become distinguishable to the environment, and the transpose fidelity drops accordingly. The channel is exactly solvable (Supplemental Material~\ref{sm:4lvl}); for the codespace $\mathcal{C}=\operatorname{span}\{|0\ket,|1\ket\}$,
\begin{equation}\label{eqn:4lvlSmallTimeFidelity}
F_{\mathrm{tr}}(t)=1-\frac{1}{24} \gamma\Delta ^2 t^3+\mathcal{O}\left(t^4\right),
\end{equation}
with the cubic dependence anticipated by the state-level argument above. The coefficient is half that of the state-level bound because the transpose map cancels the deterministic phase accumulated on the decayed branch between a decay and its detection, at the price of an additional dephasing (Supplemental Material~\ref{sm:4lvl}). In the high-fidelity limit, inverting~\cref{eqn:4lvlSmallTimeFidelity} yields
\begin{equation}
    \tau(F_0) \approx \left[{\frac{24 (1-F_0)}{\gamma  \Delta ^2}}\right]^{1/3}.
\end{equation}
The checking time therefore decreases with both the decay rate and the lifting strength: as decays become more frequent or the levels more distinguishable, recovery must be performed more often to maintain a fixed target fidelity. If either $\gamma$ or $\Delta$ vanishes, it becomes formally infinite, recovering the noiseless and perfectly degenerate limits.

The same scaling shows why repeated checking restores near-perfect recovery. Suppose the system is checked every $\tau$ over a total time $T=N\tau$, with the transpose recovery applied after each check. Each cycle then acts on the codespace as a dephasing channel with a real, positive coherence factor (Supplemental Material~\ref{sm:4lvl}), so the infidelities of successive cycles add:
$1-F_{\mathrm{tr}}^{(N)}\le N[1-F_{\mathrm{tr}}(\tau)]=\frac{1}{24}\gamma\Delta^2T\tau^2+\mathcal{O}(\tau^3)$. Here $F_{\mathrm{tr}}^{(N)}$ is the entanglement fidelity of the $N$-cycle channel $(\mathcal{R}_{\mathrm{tr}}\circ\mathcal{N}_{\tau})^{\circ N}$. Keeping the total infidelity below $\epsilon$ therefore requires, to leading order,
\begin{equation}
    \tau \lesssim \left(\frac{24\,\epsilon}{\gamma\Delta^2T}\right)^{1/2}.
\end{equation}
Checking this often, with decoding whenever a decay is detected, prevents distinguishability from building up and recovers the encoded state with fidelity approaching one.

%% ======================================================================
%% Atomic and molecular noise model
%% ======================================================================
\paragraph{Atomic and molecular systems.}
We now develop a canonical spontaneous-emission noise model for atomic and molecular angular-momentum systems, the finite-dimensional angular-momentum analogue of bosonic amplitude damping (see Refs.~\cite{chessa2021quantum,chessa2023resonanta} for multilevel damping without angular-momentum structure). It captures the universal electric-dipole selection rules and gives a platform-independent baseline for monolithic QEC in atoms and molecules. We denote the simultaneous eigenstates of $\mathbf{J}^2$ and $J_z$ by $|J,m\ket$, with $\mathbf{J}^2|J,m\ket=J(J+1)|J,m\ket$ and $J_z|J,m\ket=m|J,m\ket$.
The system, the photon bath, and their coupling are described by $H=H_{\mathcal{S}}+H_{\mathcal{B}}+H_\mathrm{int}$~\cite{scully2012quantum}, where $H_{\mathcal{S}}=\sum_{J,m}\omega_0\,\eta_J\,|J,m\ket\bra J,m|$ is the internal energy, with $\omega_0$ setting the overall scale and $\eta_J$ dimensionless, $H_{\mathcal{B}}$ is the free-field Hamiltonian of the photon bath, and $H_\mathrm{int}=-\vec{d}\cdot\vec{E}$ is the electric-dipole coupling between the dipole operator $\vec{d}$ and the field $\vec{E}$ at the emitter (Supplemental Material~\ref{sm:jumps}).

The scaling $\eta_J$, which we take to increase monotonically with $J$, is species dependent. For example, $\eta_J=J(J+1)$ gives the rigid linear rotor. Unperturbed energies depend only on $J$, so the states within each $J$ manifold are degenerate.

Using the universal Lindblad equation~\cite{nathan2020universal}, the jump operators at zero temperature are
\begin{equation}\label{eqn:jumps-atom-main-text}
L_{\delta} = \sqrt{\gamma}\sum \limits_{J\geq1,|m|\leq J}f(J) \,C_{1,\delta;J,m}^{J-1,m+\delta} \,|J-1,m+\delta\ket\bra J,m|
\end{equation}
for $\delta\in\{0,\pm1\}$. The resulting changes in the magnetic quantum number $m\mapsto m+\delta$ corresponds to the three dipole-emission polarizations (Supplemental Material~\ref{sm:jumps}). Here $C$ denotes Clebsch--Gordan coefficients and $\gamma$ is the effective decay rate. The function $f(J)$ is independent of $m$ and absorbs system-specific parameters. We set $f(J)=\sqrt{(2J+1)/(2J-1)}$, so that every state $|J,m\ket$ decays at the same total rate $\gamma$, which allows a fair comparison between codes hosted in different $J$ manifolds. We also set $\eta_J=J(J+1)$ in all simulations.

To lift the degeneracy of the decay processes, we add the term
\begin{equation}\label{eqn:atomic-non-degeneracy}
    H_\Delta=
   \Delta \sum_{J,m} J\, m \,|J,m\ket \bra J,m|.
\end{equation}
The factor of $J$ makes the transition-frequency shift depend on the source state, thereby distinguishing decay pathways. A perturbation proportional to $m$ alone would shift every transition $|J,m\ket\to|J-1,m+\delta\ket$ by $-\delta\Delta$, independent of $m$. \cref{eqn:atomic-non-degeneracy} instead gives $\Delta[m-(J-1)\delta]$. We retain the unperturbed jump operators because, for $J\Delta\ll\omega_0$, corrections to their amplitudes are only of relative order $J\Delta/\omega_0$, and the results below are independent of $\omega_0$.

%% ======================================================================
%% Noise behaviour
%% ======================================================================

\begin{figure}[t!]
    \centering
    \includegraphics[width = \columnwidth]{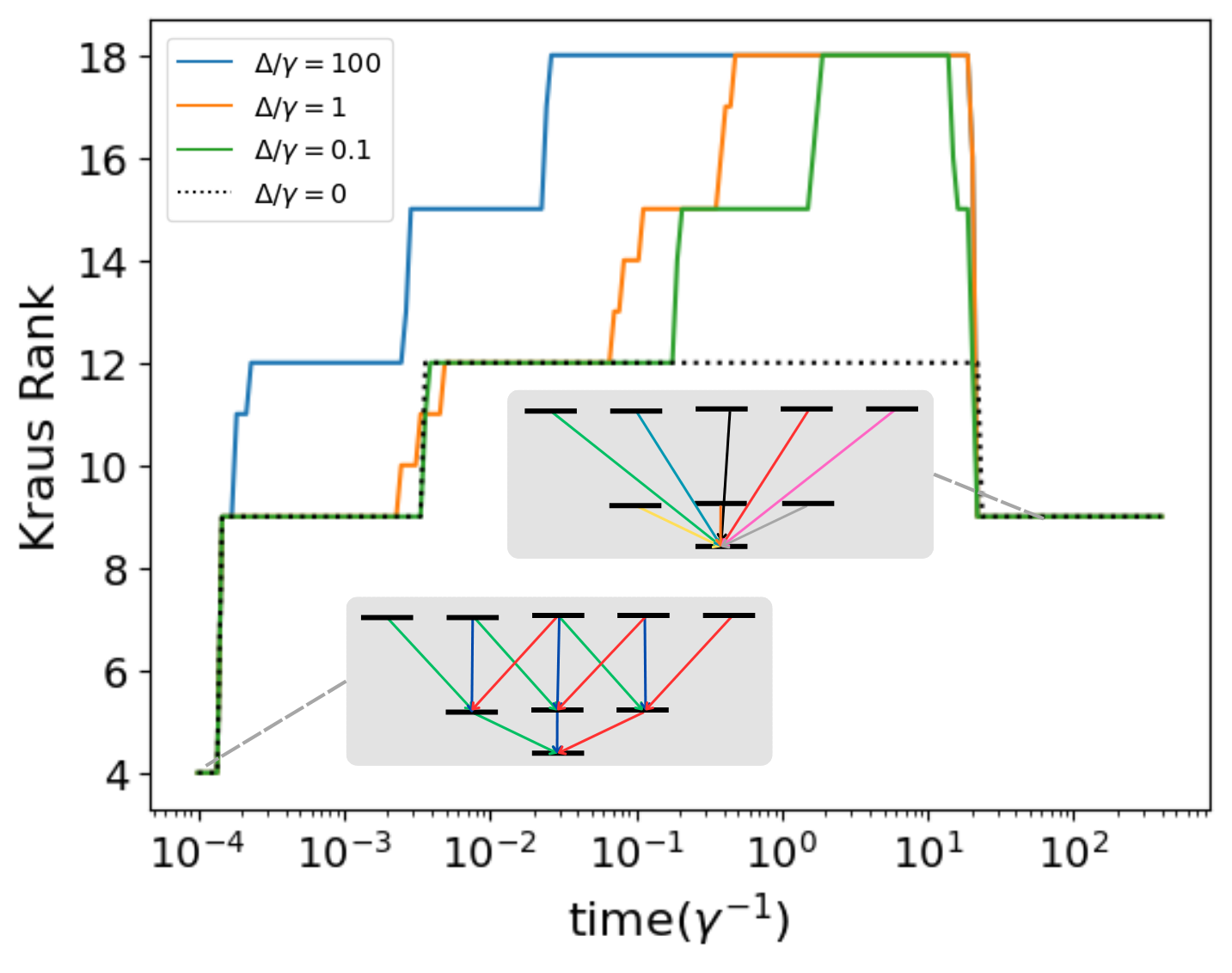}
    \caption{Effective Kraus rank versus time for the channel generated by $H_{\mathcal{S}}+H_\Delta$ and the jump operators $L_\delta$ with $J\leq2$, counting Kraus operators with weight above $10^{-8}$. The insets sketch the error channel at early and late times, with colors denoting distinct Kraus operators: early Kraus operators have support on several source states, whereas in the long-time limit each maps a single source state to the ground state $|0,0\ket$.}\label{fig:kraus_rank_linear_J}
\end{figure}
To see how the lift reshapes the channel, we take a restricted Hilbert space with $J\leq2$, which is exact at zero temperature since no transition raises $J$. We count the canonical Kraus operators $E_k$ whose weight $w_k=\Tr(E_k^\dagger E_k^{})$ exceeds $10^{-8}$, and call their number the effective Kraus rank. $w_k/d$ is the probability of error branch $k$ for a maximally mixed input in dimension $d$. For the nonzero lifting strengths in \cref{fig:kraus_rank_linear_J}, the effective rank starts at four (the three $L_\delta$ and the no-jump operator required for trace preservation~\cite{albert2018lindbladians}), rises to as many as eighteen at intermediate times, and falls back to nine in the long-time limit, when every basis state has decayed to $|0,0\ket$ and the channel is represented by the nine operators $|0,0\ket\bra J,m|$. The exact rank, counting every nonzero weight, is twelve for $\Delta=0$ and eighteen for $\Delta\neq0$ at every finite $t>0$. The effective rank records when these directions become non-negligible.

Even at $\Delta=0$ the effective rank rises to twelve at intermediate times (dotted curve), because the successive decays $J=2\to1\to0$ have distinct jump-time profiles. The additional growth for $\Delta\neq0$ is the many-level analogue of the four-level splitting. Before full decay the channel coherently sums over histories with different jump times, polarizations, and source states, and the lifting term assigns them different phases. As these phases accumulate, formerly grouped decay processes acquire weight along independent Choi directions, sooner for larger $\Delta$. Kraus rank is not a performance metric by itself, but it tracks the mechanism that degrades recovery: the environment gains access to more distinguishable error components.

%% ======================================================================
%% Code performance and fast checking
%% ======================================================================
We now ask how candidate encodings perform under this channel and whether the degradation can be mitigated. Such codes have no closed-form finite-time fidelities, so we evaluate the transpose fidelity numerically.

\begin{figure}[t!]
    \centering
    \includegraphics[width = \columnwidth]{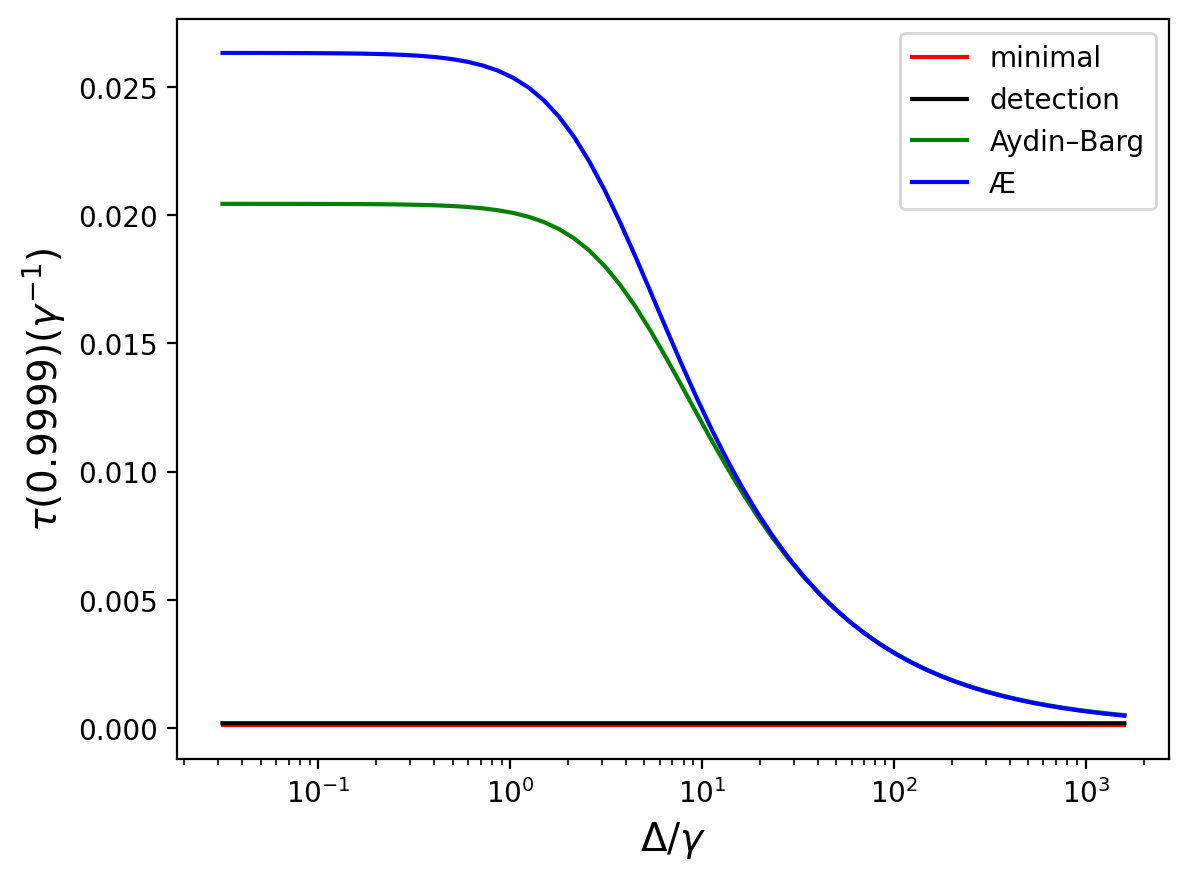}
    \caption{Checking time $\tau(F_0)$ at target fidelity $F_0=0.9999$ as a function of the degeneracy-lifting strength $\Delta$, with every state decaying at rate $\gamma$. At small $\Delta$ the checking time saturates at a code-dependent value $\tau^\star$, with $\tau^{\star \textrm{minimal}}<\tau^{\star \textrm{detection}}<\tau^{\star \textrm{Aydin--Barg}}<\tau^{\star \textrm{\AE}}$. The \AE\ curve stays above the Aydin--Barg curve until $\Delta/\gamma\approx50$, after which Aydin--Barg is marginally ahead. 
    }

\label{fig:checking_time_Delta_varying}
\end{figure}

\Cref{fig:checking_time_Delta_varying} shows the checking time of four codes, defined in the End Matter, as a function of $\Delta$. We call the single-step changes $J\to J-1$ and $m\to m+\delta$ generated by \cref{eqn:jumps-atom-main-text} rank-$1$ transitions (at zero temperature $J$ can only decrease in this model). The codes are a \emph{minimal} encoding, a \emph{detection} code that detects rank-$1$ transitions without correcting them, the absorption-emission (\AE) code~\cite{jain2024absorptionemission}, which corrects them with nonoverlapping error spaces, and the Aydin--Barg code~\cite{aydin2025class}, which corrects them with smaller separation (spacing) between the supporting states. For every code that retains part of the encoded information after a decay, the checking time decreases as distinguishability increases, consistent with the four-level analysis and the Kraus-rank growth above. The minimal code, whose codewords decay to $|0,0\ket$ by emitting photons of opposite polarization, is already fully distinguishable at $\Delta=0$, so its checking time is independent of $\Delta$. The \AE\ code performs best for $\Delta/\gamma\lesssim50$; beyond this point the two rank-$1$-correcting codes are nearly indistinguishable, with Aydin--Barg marginally ahead.
Since both codes correct the same errors, the \AE\ advantage suggests that larger separation between the supporting states improves robustness once degeneracy lifting makes the emitted photons more informative. This comparison normalizes the total decay rate of every state to $\gamma$. In real species the rates depend strongly on $J$, which may reorder the codes. The form of the lift matters as well: if the factor $J$ in \cref{eqn:atomic-non-degeneracy} is replaced by the rotational energy $J(J+1)$, the effective lift grows with the host manifold and penalizes the higher-$J$ \AE\ code, moving the crossover down to $\Delta/\gamma\approx1$ with Aydin--Barg ahead by up to 20\% thereafter.

Finally, for codes that correct rank-$1$ transitions, the degradation from degeneracy lifting can be mitigated by rapid syndrome checking. Distinguishability is produced not by the decay alone but by the time the system evolves under the degeneracy-lifted Hamiltonian before the error is identified. If the lift is calibrated, deterministic phases in both the null and detected-decay branches can be incorporated into the recovery, leaving only residual dephasing from the unknown jump time within the checking interval. An augmented $J_z$-syndrome check may provide a calibration-free alternative (Supplemental Material~\ref{sm:fast}). If a decay is detected after a short interval, the system has spent at most that interval in the lower-$J$ manifold, so the detected error approaches the ideal correctable rank-$1$ transition as the interval goes to zero.
Multi-decay events within one interval are negligible in the same limit. Continuous checking with decoding on detection therefore restores near-perfect protection even though the unchecked channel becomes increasingly distinguishable as $\Delta$ grows. We quantify how fast syndromes must be checked to restore correctability despite degeneracy lifting.

%% ======================================================================
%% Discussion
%% ======================================================================
\paragraph{Discussion.}
The which-path leakage behind these results is not specific to angular-momentum systems. Whenever a frequency shift distinguishes a decay event, the environment learns something about which state decayed, and the information--disturbance tradeoff~\cite{kretschmann2008informationdisturbance} then increases the damage that decay can cause.
Bosonic codes in weakly anharmonic oscillators face the same issue: frequency shifts make nominally degenerate decay events partially distinguishable. This is already recognized as a practical constraint in circuit-QED implementations~\cite{cai2021bosonica,brady2024advancesa,sivak2023realtimec}, where the same anharmonicity is a requirement for level addressability~\cite{girvin2014circuit}.
For Gottesman--Kitaev--Preskill (GKP) codes this contributes to the optimal envelope size: a larger mean photon number $\bar n$ better approximates the ideal code, while Kerr shifts make the occupied Fock states more distinguishable. So the logical lifetime need not improve monotonically with $\bar n$~\cite{cai2021bosonica,brady2024advancesa,sivak2023realtimec}. Approximate QEC~\cite{leung1997approximate,beny2010general} provides the language for this regime, and degeneracy lifting is one concrete and tunable way in which the KL conditions fail. Distinguishability is harmful when the resulting record depends on the logical state. Erasure conversion instead produces an error flag designed to reveal the syndrome without revealing logical information~\cite{scholl2023erasure,ma2023highfidelity}.

A second outcome of this work is the canonical angular-momentum damping model itself.
By separating universal selection-rule structure from nonuniversal reduced matrix elements, the model gives a useful baseline for comparing encodings across different atomic and molecular platforms~\cite{cornish2024quantum}.
To the best of our knowledge, no such formulation existed prior to this work. We expect it to serve atomic and molecular QEC as amplitude damping serves bosonic codes, as a common benchmark for designing codes and decoders.

Degeneracy lifting does not simply invalidate a code. The code and the lifting strength together set a checking time, the operational scale on which distinguishability accumulates, and protection survives only if syndromes are extracted faster than that. Fast checking extracts the error syndrome before appreciable which-path information accumulates in the environment. A monolithic code that relies on degenerate error processes should thus be judged together with its platform: the code, the control of the stray fields that set the lifting strength, and the syndrome-extraction speed cannot be chosen independently.
Future work could extend this viewpoint to models in which degeneracy lifting also modifies the jump operators themselves, and to continuously monitored~\cite{ahn2002continuousa,oreshkov2007continuous} or dissipatively stabilized encodings, where measurement and engineered dissipation compete on comparable timescales.

\begin{acknowledgments}
\paragraph{Acknowledgments.} 

V.V.A.\@ acknowledges NSF grants OMA-2120757 (QLCI) and CIF-2330909. A.V.G., W.C.C. and E.R.H were supported in part by NQVL:QSTD:Design:FTL and the AFOSR MURI, ARL (W911NF-24-2-0107). A.V.G. was also supported by NSF QLCI (award No.~OMA-2120757), NSF STAQ program, ONR MURI, and DoE ASCR Quantum Testbed Pathfinder program (award No.~DE-SC0024220). A.V.G.~also acknowledges support from the U.S.~Department of Energy, Office of Science, National Quantum Information Science Research Centers, Quantum Systems Accelerator (award No.~DE-SCL0000121) and from the U.S.~Department of Energy, Office of Science, Accelerated Research in Quantum Computing, Fundamental Algorithmic Research toward Quantum Utility (FAR-Qu). W.C.C.\ and E.R.H.\ acknowledge support from NSF PHY-2607831, OMA-2016245 and ARO W911NF-24-S-0004.
E.R.H. acknowledges support from NSF PHY-2409738. 
\end{acknowledgments}
\appendix
\renewcommand{\tocname}{Supplemental Material}
\clearpage
\renewcommand{\thefigure}{\arabic{figure}}
\renewcommand{\thetable}{\arabic{table}}

\onecolumngrid
\section*{End Matter}
\begin{figure}[!ht]
\centering
\begin{tikzpicture}[
  x=0.95cm,y=0.95cm,
  level/.style={black,line width=2.4pt,line cap=butt},
  vacant/.style={black!35,line width=2.4pt,line cap=butt},
  ampzero/.style={text=orange!65!black,font=\small},
  ampone/.style={text=blue!65!black,font=\small},
  jlab/.style={font=\large,anchor=east},
  code/.style={font=\large\bfseries,anchor=west},
  note/.style={font=\small,text=black!70,anchor=west}
]
\newcommand{\SJlevelbar}[3]{\draw[#3] (#1-0.38,#2) -- (#1+0.38,#2);}
\newcommand{\SJbluesphere}[2]{\shade[ball color=blue!80!cyan] (#1,#2+0.36) circle (0.19cm);}
\newcommand{\SJyellowsphere}[2]{\shade[ball color=yellow!85!orange] (#1,#2+0.36) circle (0.19cm);}

% J = 5: AE code  (y=8.2)
\foreach \m in {-5,...,5} {\SJlevelbar{\m}{7.4}{level}}
\SJbluesphere{-5}{7.4} \SJyellowsphere{-2}{7.4} \SJbluesphere{2}{7.4} \SJyellowsphere{5}{7.4}
\node[ampone, above=1pt] at (-5,7.98) {$\sqrt{2/7}$};
\node[ampzero,above=1pt] at (-2,7.98) {$\sqrt{5/7}$};
\node[ampone, above=1pt] at ( 2,7.98) {$\sqrt{5/7}$};
\node[ampzero,above=1pt] at ( 5,7.98) {$\sqrt{2/7}$};
\node[jlab] at (-6.0,7.4) {J\,=\,5};
\node[code] at (6.4,7.4) {\AE};
\node[note] at (6.4,7.0) {corrects rank-1 errors};

% J = 4: Aydin--Barg code  (y=6.4)
\foreach \m in {-4,...,4} {\SJlevelbar{\m}{5.8}{level}}
\SJyellowsphere{-4}{5.8} \SJbluesphere{-2}{5.8} \SJyellowsphere{2}{5.8} \SJbluesphere{4}{5.8}
\node[ampzero,above=1pt] at (-4,6.38) {$\sqrt{1/3}$};
\node[ampone, above=1pt] at (-2,6.38) {$\sqrt{2/3}$};
\node[ampzero,above=1pt] at ( 2,6.38) {$\sqrt{2/3}$};
\node[ampone, above=1pt] at ( 4,6.38) {$-\sqrt{1/3}$};
\node[jlab] at (-6.0,5.8) {J\,=\,4};
\node[code] at (6.4,5.8) {Aydin--Barg};
\node[note] at (6.4,5.4) {corrects rank-1 errors};

% J = 3: unoccupied in the portfolio  (y=4.6)
\foreach \m in {-3,...,3} {\SJlevelbar{\m}{4.2}{vacant}}
\node[jlab,text=black!50] at (-6.0,4.2) {J\,=\,3};

% J = 2: detection code  (y=2.8)
\foreach \m in {-2,...,2} {\SJlevelbar{\m}{2.6}{level}}
\SJyellowsphere{-2}{2.6} \SJbluesphere{0}{2.6} \SJyellowsphere{2}{2.6}
\node[ampzero,above=1pt] at (-2,3.18) {$1/\sqrt2$};
\node[ampone, above=1pt] at ( 0,3.18) {$1$};
\node[ampzero,above=1pt] at ( 2,3.18) {$1/\sqrt2$};
\node[jlab] at (-6.0,2.6) {J\,=\,2};
\node[code] at (6.4,2.6) {Detection};
\node[note] at (6.4,2.2) {detects rank-1 errors};

% J = 1: minimal code  (y=1.0)
\foreach \m in {-1,...,1} {\SJlevelbar{\m}{1.0}{level}}
\SJyellowsphere{-1}{1.0} \SJbluesphere{1}{1.0}
\node[ampzero,above=1pt] at (-1,1.58) {$1$};
\node[ampone, above=1pt] at ( 1,1.58) {$1$};
\node[jlab] at (-6.0,1.0) {J\,=\,1};
\node[code] at (6.4,1.0) {Minimal};
\node[note] at (6.4,0.6) {lowest $J$ that detects decays};

% Shared m axis
\draw[black!80,line width=0.5pt] (-5.6,0.1) -- (5.6,0.1);
\foreach \m in {-5,...,5} {\node[below=3pt] at (\m,0.1) {$\m$};}
\node[below=15pt] at (0,0.1) {$m$};
\end{tikzpicture}
\caption{Representative encodings used in the numerical comparisons. Each row of black bars is a degenerate angular-momentum manifold $J$, and horizontal position gives the magnetic quantum number $m$. Yellow and blue spheres mark the basis states supporting $|\bar0\ket$ and $|\bar1\ket$, respectively, with their nonzero amplitudes indicated. The minimal and detection encodings serve as baselines, while the \AE\ and Aydin--Barg codes both correct rank-$1$ transitions.}
\label{fig:code-portfolio-stacked}
\end{figure}
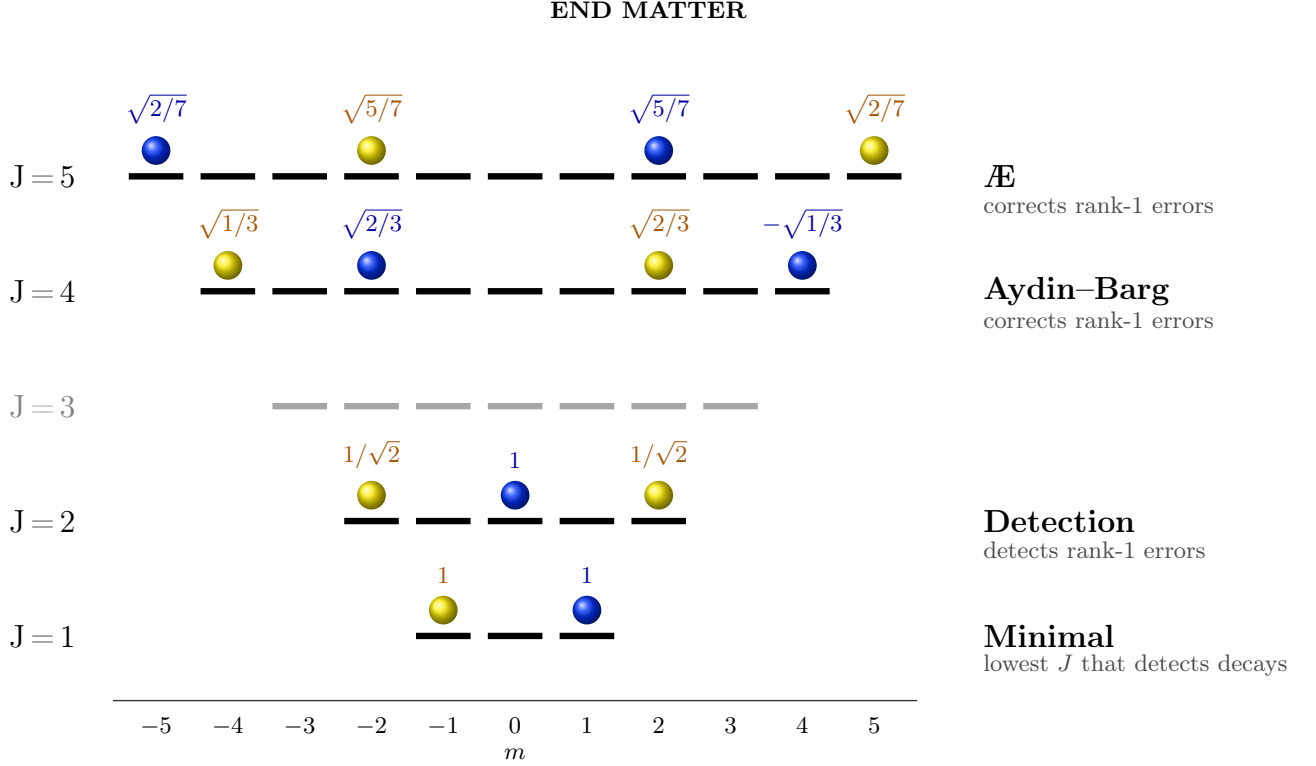
\twocolumngrid

%% ======================================================================
%% End Matter
%% ======================================================================
This section defines the representative codes used in the numerical comparisons in the main text and gives the Kraus operators of the four-level channel. \AE\ codes are the general class of codes that protect against the absorption-emission noise~\cite{jain2024absorptionemission} considered in this work. We select several code instances relevant to this noise model and compare their performance under degeneracy lifting. Our portfolio of codes, illustrated in \cref{fig:code-portfolio-stacked}, comprises:

\begin{itemize}\setlength{\itemsep}{2pt}
\item \emph{Minimal code}: $|\bar{0}\ket=|1,-1\ket$, $|\bar{1}\ket=|1,1\ket$. It is the smallest encoding supported within a single $J$ manifold and serves as a baseline. A decay is flagged, since the state leaves the $J=1$ manifold, but cannot be corrected because both codewords decay to $|0,0\ket$. 
\item \emph{Detection code}: $|\bar{0}\ket=\left(|2,-2\ket+|2,2\ket\right)/\sqrt{2}$, $|\bar{1}\ket=|2,0\ket$. Rank-$1$ transitions take any encoded state to a subspace orthogonal to the codespace, so they are detected but not corrected. Unlike for the minimal code, the error spaces of the two codewords only partially overlap, so a detected decay retains part of the encoded information, giving a longer checking time. It is an instance of the \AE\ codes of Ref.~\cite{jain2024absorptionemission}.
\item \emph{\AE\ code}: $|\bar{0}\ket=\sqrt{5/7}\,|5,-2\ket+\sqrt{2/7}\,|5,5\ket$, $|\bar{1}\ket=\sqrt{2/7}\,|5,-5\ket+\sqrt{5/7}\,|5,2\ket$. Corrects rank-$1$ transitions with nonoverlapping error spaces, so that every condition $\bra\bar{0}|E_k^\dagger E_l^{}|\bar{1}\ket=0$ is trivially satisfied for rank-$1$ errors $E_k,E_l$~\cite{jain2024absorptionemission}.
\item \emph{Aydin--Barg code}: $|\bar{0}\ket=\sqrt{1/3}\,|4,-4\ket+\sqrt{2/3}\,|4,2\ket$, $|\bar{1}\ket=\sqrt{2/3}\,|4,-2\ket-\sqrt{1/3}\,|4,4\ket$. Corrects the same rank-$1$ transitions with smaller separation between the supporting states, achieving the same code distance at lower $J$~\cite{aydin2025class}.
\end{itemize}

%% ======================================================================
%% End Matter code figure
%% ======================================================================
% Single-panel End Matter code schematic, including the unoccupied J=3 manifold.
% Style matches Fig. 1: thick black bars per |J,m> state, ball-shaded spheres on the
% states supporting the codewords, shared m axis at the bottom.
% Requires in the preamble: \usepackage{tikz} \usetikzlibrary{positioning,shadings}
%
% Yellow spheres: support of |\bar 0\rangle.   Blue spheres: support of |\bar 1\rangle.
% Labels above spheres are the nonzero codeword amplitudes.

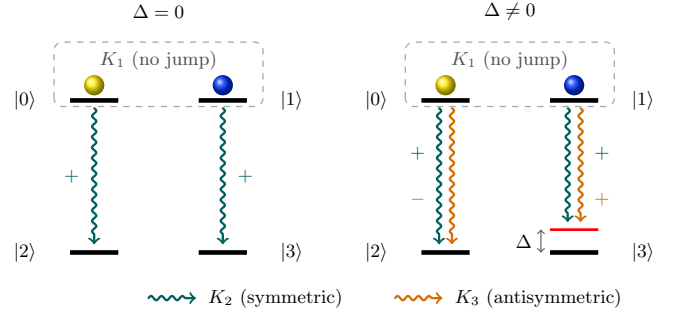
\begin{figure}[h!tbp]
\centering
\resizebox{\columnwidth}{!}{%
\begin{tikzpicture}[x=0.9cm,y=0.85cm,
  level/.style={black,line width=2.2pt,line cap=butt},
  lifted/.style={red,line width=1.3pt,line cap=butt},
  kA/.style={->,line width=1.1pt,color=teal!75!black,decorate,decoration={snake,amplitude=1pt,segment length=5pt,post length=3pt}},
  kB/.style={->,line width=1.1pt,color=orange!85!black,decorate,decoration={snake,amplitude=1pt,segment length=5pt,post length=3pt}},
  lab/.style={font=\small},
  nojump/.style={gray!70,dashed,rounded corners=4pt,line width=0.6pt}]
\newcommand{\SJlevels}{%
  \draw[level] (-0.45,3) -- (0.45,3); \node[lab,anchor=east] at (-0.95,3) {$|0\ket$};
  \draw[level] (1.95,3) -- (2.85,3); \node[lab,anchor=west] at (3.35,3) {$|1\ket$};
  \draw[level] (-0.45,0) -- (0.45,0); \node[lab,anchor=east] at (-0.95,0) {$|2\ket$};
  \draw[level] (1.95,0) -- (2.85,0); \node[lab,anchor=west] at (3.35,0) {$|3\ket$};
  \shade[ball color=yellow!85!orange] (0,3.3) circle (0.17cm);
  \shade[ball color=blue!80!cyan] (2.4,3.3) circle (0.17cm);
  \draw[nojump] (-0.75,2.88) rectangle (3.15,4.15);
  \node[lab,gray!60!black,anchor=north] at (1.2,4.12) {$K_1$ (no jump)};}
% ---------- Delta = 0 ----------
\begin{scope}
\node[font=\small\bfseries] at (1.2,4.75) {$\Delta=0$};
\SJlevels
\draw[kA] (0,2.86) -- (0,0.16);
\draw[kA] (2.4,2.86) -- (2.4,0.16);
\node[lab,teal!75!black,anchor=east] at (-0.15,1.5) {$+$};
\node[lab,teal!75!black,anchor=west] at (2.55,1.5) {$+$};
\end{scope}
% ---------- Delta != 0 ----------
\begin{scope}[xshift=5.9cm]
\node[font=\small\bfseries] at (1.2,4.75) {$\Delta\neq0$};
\SJlevels
\draw[lifted] (1.95,0.45) -- (2.85,0.45);
\draw[<->,black!70,line width=0.5pt] (1.78,0) -- (1.78,0.45); \node[lab,anchor=east] at (1.74,0.22) {$\Delta$};
\draw[kA] (-0.12,2.86) -- (-0.12,0.16);
\draw[kB] ( 0.12,2.86) -- ( 0.12,0.16);
\draw[kA] (2.28,2.86) -- (2.28,0.6);
\draw[kB] (2.52,2.86) -- (2.52,0.6);
\node[lab,teal!75!black,anchor=east] at (-0.25,1.95) {$+$};
\node[lab,orange!85!black,anchor=east] at (-0.25,1.05) {$-$};
\node[lab,teal!75!black,anchor=west] at (2.65,1.95) {$+$};
\node[lab,orange!85!black,anchor=west] at (2.65,1.05) {$+$};
\end{scope}
% ---------- legend ----------
\draw[kA] (1.0,-0.9) -- (1.9,-0.9); \node[lab,anchor=west] at (2.0,-0.9) {$K_2$ (symmetric)};
\draw[kB] (5.6,-0.9) -- (6.5,-0.9); \node[lab,anchor=west] at (6.6,-0.9) {$K_3$ (antisymmetric)};
\end{tikzpicture}%
}
\caption{Onset of the third Kraus operator in the four-level model. At $\Delta=0$ the channel has two Kraus operators, the no-jump operator $K_1$ and a single decay operator $K_2$ that acts identically on the two decay paths. For $\Delta\neq0$ the decay branch splits into the symmetric and antisymmetric combinations $K_2$ and $K_3$ of the two paths, \cref{eqn:kraus-K2,eqn:kraus-K3}. The signs indicate the relative sign of the two paths in each operator.}
\label{fig:kraus3-onset}
\end{figure}

\paragraph{Onset of the third Kraus operator.} We return to the four-level model of \cref{eqn:4-level-hamiltonian} to exhibit how a degeneracy lift changes the noise channel itself. The Kraus operators of the channel generated by $H_\mathcal{S}+H_\Delta$ and the jump operator $\sqrt{\gamma}\left(|2\ket\bra0|+|3\ket\bra1|\right)$ are
\begin{eqnarray}
K_{1}   & = & e^{-i\omega_{0}t}e^{-\frac{\gamma t}{2}+i\Delta t}\Pi_{01}+e^{i\Delta t}\Pi_{2}+\Pi_{3},\label{eqn:kraus-K1}\\
K_{2} & = & e^{-\frac{\gamma t}{2}}\sqrt{\frac{f_+(t)}{2}}\,\Biggl[\frac{\left|\Omega\left(e^{t\Omega^\star}-1\right)\right|}{\Omega\left(e^{t\Omega^\star}-1\right)}|2\ket\bra0|\nonumber\\*
& & {}+|3\ket\bra1|\Biggr],\label{eqn:kraus-K2}\\
K_{3} & = & e^{-\frac{\gamma t}{2}}\sqrt{-\frac{f_-(t)}{2}}\,\Biggl[-\frac{\left|\Omega\left(e^{t\Omega^\star}-1\right)\right|}{\Omega\left(e^{t\Omega^\star}-1\right)}|2\ket\bra0|\nonumber\\*
& & {}+|3\ket\bra1|\Biggr],\label{eqn:kraus-K3}
\end{eqnarray}                                         
where $\Pi_X = \sum_{i\in \{X\}}|i\ket \bra i|$ is the projector onto the states in $X$, $\ensuremath{\Omega=\gamma+i\Delta}\textrm{, and }\ensuremath{f_{\pm}(t):=\frac{\gamma}{|\Omega|}|e^{\Omega t}-1|\pm\left(e^{\gamma t}-1\right)}$, with $f_-(t)\le0$, vanishing only at $\Delta=0$. In the degenerate limit ($\Delta=0$), one has $f_-(t)\equiv 0$. In this regime, the Kraus operator $K_3$ vanishes, and the remaining operators reduce to the standard amplitude-damping form
\begin{eqnarray}
K_{1}|_{\Delta=0}  = & e^{-i\omega_{0}t}\sqrt{e^{-\gamma t}}\Pi_{01}+\Pi_{23},\\
K_{2}|_{\Delta=0}  = & \sqrt{1-e^{-\gamma t}}\left(|2\ket\bra 0| + |3\ket \bra 1|\right).
\end{eqnarray} 
For nonzero $\Delta$, however, the Kraus rank increases from $2$ to $3$ for any $t>0$. The appearance of $K_3$ is the signature of the channel splitting discussed in the main text: it captures the additional distinguishability created by lifting the degeneracy and therefore marks the onset of imperfect recoverability. As $\Delta$ grows, this additional Kraus direction generally carries more weight, reflecting the increasing distinguishability between the two decay paths. Explicitly, the weights of the two decay operators are $w_{2,3}=\Tr(K_{2,3}^{\dagger}K_{2,3}^{})=p_{\mathrm{decay}}\left(1\pm|\chi_{\mathrm{d}}(t)|\right)$, where $\chi_{\mathrm{d}}(t)=\gamma\left(e^{i\Delta t}-e^{-\gamma t}\right)/\left[(\gamma+i\Delta)(1-e^{-\gamma t})\right]$ is the coherence retained on the decayed branch (Supplemental Material~\ref{sm:4lvl}), so $K_3$ carries the fraction $(1-|\chi_{\mathrm{d}}|)/2$ of the decay branch, which grows as $\Delta^2t^2/48$ at short times and saturates at $\frac{1}{2}\left(1-\gamma/\sqrt{\gamma^2+\Delta^2}\right)$. Here $|\chi_{\mathrm{d}}|$ can be read as the overlap of the photon states emitted along the two decay paths, and $K_2$ and $K_3$ are their symmetric and antisymmetric combinations (\cref{fig:kraus3-onset}).

\onecolumngrid
%% ======================================================================
%% Supplemental Material
%% ======================================================================
\newpage
\section*{Supplemental Material}

%% ======================================================================
%% S1: Four-Level System
%% ======================================================================
\appsection{Four-Level System}\label{sm:4lvl}
This section provides the explicit calculations for the four-level example discussed in the main text. The system is governed by the Hamiltonian 
    $H_\mathcal{S} = \omega_0\left (|0\ket \bra0| + |1\ket\bra 1|\right )$
with jump operator
    $\sqrt{\gamma}\left(|2\ket\bra 0| + |3\ket\bra 1|\right)$
and lifting term
    $H_\Delta=\Delta |3\ket \bra 3|$.

\subsection{State-Level Fidelity}

We first derive the state-level fidelity quoted in the main text. For an initial code state $|\psi\ket=a|0\ket+b|1\ket$, the evolved density matrix is
\begin{eqnarray}
\rho(t)  = &  e^{-\gamma t}|\psi\ket\bra\psi|
+\left[|a|^2|2\ket\bra2|+|b|^2|3\ket\bra3|\right]\left(1-e^{-\gamma t}\right) \nonumber \\
& + \left[\frac{\gamma a b^\star\left(e^{i\Delta t}-e^{-\gamma t}\right)}{\gamma+i\Delta}|2\ket\bra3|+ \textrm{h.c.}\right].
\end{eqnarray}
The null outcome has probability $p_{\mathrm{null}}=e^{-\gamma t}$ and returns the initial code state after normalization. The decay outcome has probability $p_{\mathrm{decay}}=1-e^{-\gamma t}$. After applying $\mathcal{R}_4$, its normalized decoded state is
\begin{eqnarray}
\rho_{\mathrm{d}}(t)&=&|a|^2|0\ket\bra0|+|b|^2|1\ket\bra1|
+\left[a b^\star\chi_{\mathrm{d}}(t)|0\ket\bra1|+\textrm{h.c.}\right],
\end{eqnarray}
where
\begin{equation}\label{eqn:chi-d}
\chi_{\mathrm{d}}(t)=
\frac{\gamma\left(e^{i\Delta t}-e^{-\gamma t}\right)}
{(\gamma+i\Delta)\left(1-e^{-\gamma t}\right)}.
\end{equation}
Thus the branch-averaged decoded state fidelity is
\begin{eqnarray}
F_\psi(t)&=&p_{\mathrm{null}}+p_{\mathrm{decay}}\bra\psi|\rho_{\mathrm{d}}(t)|\psi\ket\nonumber\\
&=&1-2|a|^2|b|^2\left(1-e^{-\gamma t}\right)\left(1-\mathrm{Re}\,\chi_{\mathrm{d}}(t)\right).
\end{eqnarray}
Expanding at short times gives
\begin{equation}
1-F_\psi(t)=\frac{|a|^2|b|^2}{3}\gamma\Delta^2t^3+\mathcal{O}(t^4).
\end{equation}

\subsection{Transpose Fidelity}

The Kraus operators $K_k$ of this channel, and the onset of $K_3$ for $\Delta\neq0$, are given in the End Matter, \crefrange{eqn:kraus-K1}{eqn:kraus-K3}. They allow the transpose fidelity, \cref{eqn:transpose-fidelity} of the main text, to be evaluated exactly. For a channel with Kraus operators $\{E_l\}$ and a code with basis $\{|\bar\nu\ket\}$ of dimension $d_{\mathrm{L}}$, it reads
\begin{equation}
F_{\mathrm{tr}}=\frac{1}{d_{\mathrm{L}}^2}\left\lVert\Tr_{\mathrm{L}}\sqrt{M}\right\rVert_{\mathrm{F}}^2,\qquad M_{[\mu,l],[\nu,k]}=\bra\bar\mu|E_l^\dagger E_k^{}|\bar\nu\ket,
\end{equation}
where $M$ is the QEC matrix, $(\Tr_{\mathrm{L}} B)_{l,k}=\sum_\mu B_{[\mu,l],[\mu,k]}$ is the partial trace over the logical index, and $\lVert B\rVert_{\mathrm{F}}^2=\Tr(B^\dagger B)$ is the squared Frobenius norm. For the codespace $\mathcal{C}=\operatorname{span}\{|0\ket,|1\ket\}$, one obtains
\begin{equation}\label{eqn:fidelity-4level-system}
    F_{\mathrm{tr}}(t)=\frac{\left(\gamma ^2+\Delta ^2\right) \sinh (\gamma  t)+\gamma ^2 [\cosh (\gamma  t)-\cos (\Delta  t)]}{\left(\gamma ^2+\Delta ^2\right) \left(e^{\gamma  t}-1\right)}.
\end{equation}
Equivalently, $F_{\mathrm{tr}}(t)=1-\frac{p_{\mathrm{decay}}}{2}\left(1-|\chi_{\mathrm{d}}(t)|^2\right)$. Expanding this expression at short times gives the cubic decay quoted in the main text. At long times, the fidelity saturates to
\begin{equation}
    F_{\mathrm{tr}}(t\rightarrow\infty)=\frac{1}{2}\left(1+\frac{\gamma^2}{\gamma^2+\Delta^2}\right).
\end{equation}
Thus the degenerate limit $\Delta=0$ remains perfectly recoverable, while for $\Delta\gg\gamma$ the late-time fidelity approaches $1/2$, as expected when the two decay paths become fully distinguishable.

\subsection{Comparison of Recoveries and Fidelity Measures}
The state-level fidelity $F_\psi$ and the transpose fidelity $F_{\mathrm{tr}}$ differ in both the fidelity measure and the recovery. Every recovery considered here returns the null branch unchanged, so the recoveries differ only in the coherence factor $\lambda$ that the logical qubit retains on the decayed branch, and $F_{\mathrm{e}}=1-\frac{p_{\mathrm{decay}}}{2}(1-\mathrm{Re}\,\lambda)$. For $\mathcal{R}_4$, $\lambda=\chi_{\mathrm{d}}$ and the entanglement infidelity then coincides with the worst-case state infidelity, $1-F_{\mathrm{e}}=\max_\psi(1-F_\psi)=\frac{1}{12}\gamma\Delta^2t^3$. The state infidelity averaged over Haar-random pure logical states is $\frac{2}{3}$ of this value, $\frac{1}{18}\gamma\Delta^2t^3$, in accordance with $\bar F=(2F_{\mathrm{e}}+1)/3$~\cite{horodecki1999general,nielsen2002simple}. The transpose map can be written explicitly: with $\mathcal{N}(\Pi_{01})=e^{-\gamma t}\Pi_{01}+p_{\mathrm{decay}}\Pi_{23}$, and $\Pi_X$ and $K_k$ the projectors and the channel Kraus operators of the End Matter, its Kraus operators $\Pi_{01}K_k^\dagger\mathcal{N}(\Pi_{01})^{-1/2}$~\cite{petz1986sufficient,petz1988sufficiency,barnum2002reversing,ng2010simple} are
\begin{equation}
R_1=\Pi_{01},\qquad R_2=\sqrt{\tfrac{1+|\chi_{\mathrm{d}}|}{2}}\,D_+,\qquad R_3=\sqrt{\tfrac{1-|\chi_{\mathrm{d}}|}{2}}\,Z D_+,\qquad D_+=e^{-i\arg\chi_{\mathrm{d}}}|0\ket\bra2|+|1\ket\bra3|,
\end{equation}
with $Z=|0\ket\bra0|-|1\ket\bra1|$. Each $R_k$ is defined up to a phase. The transpose map is therefore $\mathcal{R}_4$ with the rotation $e^{-i\arg\chi_{\mathrm{d}}}$ built in, which cancels the deterministic phase $\arg\chi_{\mathrm{d}}\simeq\Delta t/2$ accumulated on $|3\ket$ between the decay and its detection, followed by a $Z$ dephasing with probability $(1-|\chi_{\mathrm{d}}|)/2$. The net coherence factor is $\lambda=|\chi_{\mathrm{d}}|^2$, which reproduces \cref{eqn:fidelity-4level-system}. The phase-corrected decoder $D_+$ alone gives $\lambda=|\chi_{\mathrm{d}}|$ and $1-F_{\mathrm{e}}=\frac{1}{48}\gamma\Delta^2t^3$, and is in fact optimal at all times: the null and decayed branches are orthogonal in the output, and on the decayed branch the logical channel is a rotation followed by a Pauli-covariant dephasing. Since $F_{\mathrm{e}}(\mathcal{R}\circ U)+F_{\mathrm{e}}(\mathcal{R}\circ ZU)\le1$ for any recovery $\mathcal{R}$, no recovery can retain a coherence factor larger than $|\chi_{\mathrm{d}}|$. Hence $F_{\mathrm{opt}}=1-\frac{p_{\mathrm{decay}}}{2}(1-|\chi_{\mathrm{d}}|)$ and $1-F_{\mathrm{tr}}=(1+|\chi_{\mathrm{d}}|)(1-F_{\mathrm{opt}})$ exactly, which saturates the bound $1-F_{\mathrm{opt}}\ge\frac{1}{2}(1-F_{\mathrm{tr}})$~\cite{barnum2002reversing,zheng2024nearoptimal} as $t\to0$. In terms of the Kraus weights $w_k=\Tr(K_k^{\dagger}K_k^{})$, $1-F_{\mathrm{opt}}=w_3/2$ and $1-F_{\mathrm{tr}}=w_2w_3/(2p_{\mathrm{decay}})$: the infidelity of the optimal recovery is exactly half the weight of the third Kraus operator. At short times $\mathrm{Re}\,\chi_{\mathrm{d}}\le|\chi_{\mathrm{d}}|^2\le|\chi_{\mathrm{d}}|$, so the transpose map halves the leading infidelity of $\mathcal{R}_4$ and lies within its guaranteed factor of two of the optimum. At later times, whenever $\arg\chi_{\mathrm{d}}$ passes through a multiple of $2\pi$, $\mathrm{Re}\,\chi_{\mathrm{d}}$ exceeds $|\chi_{\mathrm{d}}|^2$ and $\mathcal{R}_4$ transiently outperforms the transpose map. These results are collected in Table~\ref{tab:4lvl-fidelities}.
\begin{table}[h]
\centering
\begin{tabular}{llll}
\hline\hline
Quantity & Recovery & Fidelity measure & $(1-F)/(\gamma\Delta^2t^3)$ \\
\hline
$\min_\psi F_\psi$ & $\mathcal{R}_4$ & worst-case state fidelity & $1/12$ \\
$F_{\mathrm{e}}$ & $\mathcal{R}_4$ & entanglement fidelity & $1/12$ \\
$\bar F$ & $\mathcal{R}_4$ & state fidelity averaged over Haar-random logical states & $1/18$ \\
$F_{\mathrm{tr}}$ & transpose & entanglement fidelity & $1/24$ \\
$F_{\mathrm{e}}$ & phase-corrected $\mathcal{R}_4$ & entanglement fidelity & $1/48$ \\
\hline\hline
\end{tabular}
\caption{Leading short-time infidelity of the four-level code, $\mathcal{C}=\operatorname{span}\{|0\ket,|1\ket\}$, for different recoveries and fidelity measures. The last row is optimal over all recoveries at every $t$.}
\label{tab:4lvl-fidelities}
\end{table}

%% ======================================================================
%% S2: Universal Lindblad Equation
%% ======================================================================
\appsection{Universal Lindblad Equation}\label{sm:ule}
In this section we summarize the ingredients of the universal Lindblad equation (ULE)~\cite{nathan2020universal} needed to derive the jump operators used in the main text. For a system $\mathcal{S}$ and bath $\mathcal{B}$, consider the Hamiltonian 
\begin{equation}
   H = H_\mathcal{S}+H_{\mathcal{B}}+H_\mathrm{int},
\end{equation}
where $H_\mathcal{S}$ and $H_\mathcal{B}$ denote Hamiltonians for the system and the bath, respectively, and $H_\mathrm{int}$ of the form
\begin{equation}
    H_\mathrm{int} = \sqrt{\gamma} \sum_\alpha X_\alpha B_\alpha
\end{equation}
contains terms coupling the system with the bath. For all $\alpha$, $X_\alpha$ and $B_\alpha$ are Hermitian operators on the system $\mathcal{S}$ and bath $\mathcal{B}$, respectively, while $\gamma$ parametrizes the coupling strength between the two. The main idea is that, when the correlation times of the bath observables $\{B_\alpha\}$ are much shorter than the inverse rate of bath-induced evolution of the system, the system dynamics are effectively Markovian. Hence, Markovianity is independent of the details of the system, and only depends on the bath. In this regime, one obtains the \emph{universal Lindblad equation} (ULE)~\cite{nathan2020universal}
\begin{equation}
    \partial_{t}\rho(t)=-i[H_{\mathcal{S}}(t)+\Lambda(t),\rho(t)]+\sum_{\lambda}\left(L_{\lambda}(t)\rho(t)L_{\lambda}^{\dagger}(t)-\frac{1}{2}\{L_{\lambda}^{\dagger}(t)L_{\lambda}(t),\rho(t)\}\right)+\xi(t).
\end{equation}
$\Lambda(t)$ is the Lamb shift, which can be viewed as a correction to the system Hamiltonian induced by the coupling term, $L_\lambda$ are the jump operators describing the dissipative dynamics, and $\xi$ is a correction term accounting for deviations from exact Markovianity.
Throughout this section, tildes denote interaction-picture operators. For any time-local system operator $O(t)$, including $L_\lambda(t)$ and $\Lambda(t)$,
\begin{equation}
    \tilde{O}(t)=U_{\mathcal{S}}^{\dagger}(t)O(t)U_{\mathcal{S}}(t),
    \qquad
    O(t)=U_{\mathcal{S}}(t)\tilde{O}(t)U_{\mathcal{S}}^{\dagger}(t),
\label{eqn:interaction-picture}\end{equation}
where $U_{\mathcal{S}}(t)=\mathcal{T}e^{-i\int_{0}^{t}ds\,H_{\mathcal{S}}(s)}$.

To derive the jump operators, one computes a small set of bath correlation functions, which then determine the dissipative channel acting on $\mathcal{S}$. We first write the matrix elements of the bath correlation function $\mathcal{J}(t)$ and its Fourier transform $\mathcal{J}(\omega)$ as
\begin{eqnarray}
    \mathcal{J}_{\alpha \beta}(t-s) &=& \Tr_\mathcal{B}\left[{\tilde{B}_\alpha(t)\tilde{B}_\beta(s)\rho_\mathcal{B}}\right], \label{eqn:bath-correlation-t}\\
    \mathcal{J}_{\alpha \beta}(\omega) &=& \frac{1}{2\pi}\int_{-\infty}^{\infty}dt \mathcal{J}_{\alpha \beta}(t) e^{i\omega t},\label{eqn:bath-correlation-w}
\end{eqnarray}
where $\tilde{B}_\alpha(t):=e^{iH_{\mathcal{B}}t}B_\alpha e^{-iH_\mathcal{B}t}$ and $\rho_{\mathcal{B}}$ is a steady state of the bath. One then computes the matrix-valued jump correlators
\begin{eqnarray}
    \textbf{g}(\omega) &=& \sqrt{\boldsymbol{\mathcal{J}}(\omega)/2\pi},\label{eqn:jump-correlation-w} \\
    \textbf{g}(t) &=& \int_{-\infty}^{\infty}d\omega \textbf{g}(\omega)e^{-i\omega t}.\label{eqn:jump-correlation-t}
\end{eqnarray}
The validity of the ULE is controlled by two parameters, $\Gamma$ and $\tau_{\rm B}$, which can be interpreted as an upper bound on the rate of bath-induced evolution and a measure of the characteristic bath-correlation time, respectively. This follows the notation of Ref.~\cite{nathan2020universal} for $\Gamma$, while we write the bath correlation time as $\tau_{\rm B}$ to distinguish it from the checking time used in the main text. They are computed from the jump correlators as
\begin{eqnarray}
    \Gamma &=& 4\gamma\left[\int_{-\infty}^{\infty} dt \lVert \textbf{g}(t)\rVert_{2,1}\right]^2, \label{eqn:timescales-ule-interaction}\\
    \tau_{\rm B} &=&\frac{\int_{-\infty}^{\infty} dt\,\lVert \textbf{g}(t)t\rVert_{2,1}} {\int_{-\infty}^{\infty} dt\lVert \textbf{g}(t)\rVert_{2,1}},\label{eqn:timescales-ule-bath}
\end{eqnarray}
where $\lVert \textbf{M} \rVert _{2,1}:=\sum_b\left(\sum_\alpha |M_{\alpha b}|^2\right)^{\frac{1}{2}}$ is the $L_{2,1}$ norm of the matrix $\textbf{M}$. The system is effectively Markovian when $\Gamma\tau_{\rm B} \ll 1$ because $\lVert\xi(t)\rVert\leq 2\Gamma^2 \tau_{\rm B}$ for all times, where $\lVert\cdot\rVert$ denotes the spectral norm, so the correction to Markovian dynamics is of order $\Gamma^2 \tau_{\rm B}$. In the interaction picture, the jump operators and Lamb shift are
\begin{eqnarray}
    \tilde{L}_{\lambda}(t)&=&\sqrt{\gamma}\sum_{\alpha}\int_{-\infty}^{\infty}ds\,g_{\lambda\alpha}(t-s)\tilde{X}_{\alpha}(s), \label{eqn:jumps-interaction}\\
    \tilde{\Lambda}(t)&=&\frac{\gamma}{2i}\int_{-\infty}^{\infty} ds \int_{-\infty}^{\infty} ds'
\sum_{\alpha\beta}\,\tilde{X}_{\alpha}(s)\tilde{X}_{\beta}(s')\phi_{\alpha\beta}(s-t,s'-t),\label{eqn:lamb-interaction}
\end{eqnarray}
where $\tilde{X}_\alpha(s)=U_{\mathcal{S}}^{\dagger}(s)X_{\alpha}U_{\mathcal{S}}(s)$ is the interaction-picture operator of \cref{eqn:interaction-picture}. The matrix elements $\phi_{\alpha \beta}$ are defined by the matrix $\boldsymbol{\phi}(t,s)=\textbf{g}(t)\textbf{g}(-s)\operatorname{sgn}(t-s)$. 

When the Hamiltonian is time independent, the jump operators and Lamb shift reduce to
\begin{eqnarray}
L_{\lambda}&=&2\pi\sqrt{\gamma}\sum_{m,n,\alpha}g_{\lambda\alpha}\left(E_n - E_m\right)X_{mn}^{(\alpha)}|m\ket\bra n|,\label{eqn:jumps-time-independent} \\
    \Lambda&=&\sum_{l,m,n,\alpha,\beta} f_{\alpha\beta}\left(E_m-E_l,E_n-E_l\right)X^{(\alpha)}_{ml}X^{(\beta)}_{ln}|m\ket\bra n|\label{eqn:lamb-time-independent},
\end{eqnarray}
    where $\{|n\ket\}$ and $\{E_n\}$ denote the eigenstates and the corresponding eigenenergies of the system Hamiltonian $H_\mathcal{S}$. The matrix elements of $X_\alpha$ are denoted by $X_{mn}^{(\alpha)}=\bra m |X_\alpha |n \ket$, and $\textbf{f}$ is defined as $\mathbf{f}(E_1,E_2)=2\pi\,\mathcal{P}\!\int_{-\infty}^{\infty} d\omega\,
\frac{\mathbf{g}(\omega-E_1)\,\mathbf{g}(\omega+E_2)}{\omega},
$
with $\mathcal{P}$ denoting the Cauchy principal-value integral. Thus, one can derive
the Lindblad equation for Markovian evolution of any system, given the Hamiltonian governing its interactions with the bath. For the isotropic photon bath of the atomic model in the main text, the bath correlation matrix is proportional to $\delta_{\alpha\beta}$, so $\Lambda$ contracts the dipole operator with itself and is a scalar under rotations. By the Wigner--Eckart theorem it is then proportional to the identity within each $J$ manifold, so it renormalizes the level scaling $\eta_J$ but cannot lift the $m$ degeneracy, and we absorb it into $H_\mathcal{S}$.

%% ======================================================================
%% S3: Derivation of the Canonical Atomic and Molecular Spontaneous-Emission Model
%% ======================================================================
\appsection{Derivation of the Atomic and Molecular Noise Model}\label{sm:jumps}
This section derives the canonical angular-momentum spontaneous-emission model used in the main text. Starting from the standard electric-dipole coupling between the system and the electromagnetic field, and applying the usual weak-coupling and Born--Markov assumptions, we show that an isotropic photon bath produces three spherical jump channels corresponding to the three dipole polarizations. The angular dependence of these jumps is fixed by Clebsch--Gordan coefficients, while nonuniversal details such as reduced matrix elements, radial factors, density-of-states factors, field-normalization constants, and additional species-specific structure are absorbed into effective parameters such as $\gamma$ and $f(J)$. Thus the model is intended as a canonical effective spontaneous-emission channel for atomic and molecular angular-momentum systems, rather than a species-specific microscopic description.

We begin from the microscopic Hamiltonian
\begin{equation}\label{eqn:atomic-hamiltonian_appx}
H=H_{\mathcal{S}}+H_{\mathcal{B}}+H_\mathrm{int},
\end{equation}
where
\begin{itemize}
\item $H_{\mathcal{S}}=\sum_{J}\sum_{m=-J}^{J}\omega_0 \,\eta_{J} \, |J,m\ket \bra J,m|$ is the unperturbed system Hamiltonian. Here $|J,m\ket$ is a system angular-momentum eigenstate with total angular momentum quantum number $J$ and magnetic quantum number $m$, $\omega_0$ sets the overall energy scale, and $\eta_J$ is a dimensionless scaling factor characterizing the energy levels as a function of $J$.
\item $H_{\mathcal{B}}=\sum_{\vec{k},\epsilon} c|\vec{k}|a_{\vec{k},\epsilon}^{\dagger}a_{\vec{k},\epsilon}$ is the photon bath Hamiltonian. Here $\vec{k}\in\mathbb{R}^3$ is the photon wave vector, $\epsilon(\vec{k})$ is a transverse photon polarization vector; for each $\vec{k}$ the label $\epsilon$ takes two values, and the two polarization vectors together with $\hat{k}=\vec{k}/|\vec{k}|$ form an orthonormal basis, and $a_{\vec{k},\epsilon}^{\dagger}$ creates a photon in the mode $(\vec{k},\epsilon)$ with frequency $\omega=c|\vec{k}|$.
\item The dipole operator is written as $\vec d=d_0\vec X$, where $\vec X=(X_x,X_y,X_z)$ is a dimensionless rank-$1$ vector operator acting on the system Hilbert space. Thus $X_\alpha$ is a system operator whose matrix elements determine the allowed transitions. 
The electric field at the position $\vec{r}_0$ of the system has the mode expansion~\cite{scully2012quantum}
\begin{equation}
\vec E=
\sum_{\vec{k},\epsilon}
\left[
\mathcal{E}_{\omega}\,\epsilon(\vec{k})a_{\vec{k},\epsilon}
+\mathcal{E}_{\omega}^{\star}\epsilon^{\star}(\vec{k})a_{\vec{k},\epsilon}^{\dagger}
\right],
\end{equation}
where, for box-normalized modes in units with $\hbar=1$, $\mathcal{E}_{\omega}=i\sqrt{\omega/(2\varepsilon_0 V)}$  
up to a phase convention, with $\varepsilon_0$ the vacuum permittivity and $V$ the quantization volume.
The dipole approximation consists of evaluating the field at $\vec{r}_0$, neglecting its variation across the emitter, so that $H_\mathrm{int}=-\vec{d}\cdot\vec{E}(\vec{r}_0)$. Defining $\kappa(\omega)=d_0\mathcal{E}_{\omega}$, and allowing this effective coupling to include the chosen field-normalization convention, gives
\begin{equation}\label{eqn:dipole-coupling-appx}
H_\mathrm{int}=-\vec d\cdot\vec E
=-\sum_{\vec{k},\epsilon,\alpha}
\left[
X_{\alpha}\left(\hat{\mathbf e}_{\alpha}\cdot\epsilon(\vec{k})\right)
\kappa^{\star}(\omega)a_{\vec{k},\epsilon}^{\dagger}+h.c.
\right],
\end{equation}
where $\alpha\in\{x,y,z\}$ and $\hat{\mathbf e}_{\alpha}$ is the Cartesian unit vector in the $\alpha$ direction.
\end{itemize}
The jump operators for the noise channel follow from the Hamiltonian description in~\cref{eqn:atomic-hamiltonian_appx}.

In the notation of the ULE setup of the previous section, the system-bath operators are
\begin{itemize}
    \item $B_\alpha={-d_0E_\alpha/\sqrt{\gamma}}$ for $\alpha\in \{x,y,z\}$,
    \item $X_\alpha$ for $\alpha \in \{x,y,z\}$.
\end{itemize}
The derivation is kept more general than the main text by first treating a bath at finite temperature $1/(k_{\mathrm{B}}\beta)$, where $k_{\mathrm{B}}$ is the Boltzmann constant. The steady state of such a bath is the thermal state $\rho_\mathcal{B}=\frac{e^{-\beta H_{\mathcal{B}}}}{\Tr[e^{-\beta H_\mathcal{B}}]}$. The relevant bath correlation functions are $\mathcal{J}_{\alpha \beta}(t) = \Tr_\mathcal{B}\left[{\tilde{B}_\alpha(t)\tilde{B}_\beta(0)\rho_\mathcal{B}}\right]$.

With the photon degrees of freedom $\vec{k},\epsilon$ denoted by $\lambda$, the photon-number basis states $|n_{\lambda_1},n_{\lambda_2},\ldots\ket$ denoted by $|\{n_\lambda\}\ket$, and the partition function defined as $Z(\beta):=\Tr(e^{-\beta H_\mathcal{B}})=\prod_{\lambda}\frac{1}{1-e^{-\beta\omega_{\lambda}}}$, the diagonal correlator is
\begin{eqnarray}
\mathcal{J}_{\alpha\alpha}(t) & = & \frac{1}{\Tr[e^{-\beta H_{\mathcal{B}}}]}\sum_{\{n_{\lambda}\}}\bra\{n_{\lambda}\}|\int d^3k_{1}d^3k_{2}\sum_{\epsilon_{1},\epsilon_{2}}\left[\hat{\mathbf e}_{\alpha}\cdot\epsilon_{1}(\vec{k}_{1})\right]\left[\hat{\mathbf e}_{\alpha}\cdot\epsilon_{2}(\vec{k}_{2})\right]\nonumber\\
& &\times\left(\kappa^{\star}(\omega_{1})a_{\vec{k}_{1},\epsilon_{1}}^{\dagger}e^{i\omega_{1}t}+\kappa(\omega_{1})a_{\vec{k}_{1},\epsilon_{1}}e^{-i\omega_{1}t}\right)
\left(\kappa^{\star}(\omega_{2})a_{\vec{k}_{2},\epsilon_{2}}^{\dagger}+\kappa(\omega_{2})a_{\vec{k}_{2},\epsilon_{2}}\right)e^{-\beta H_{\mathcal{B}}}|\{n_{\lambda}\}\ket\\
 & = & \frac{1}{Z(\beta)}\sum_{\{n_{\lambda}\}}\bra\{n_{\lambda}\}|\int d^3k_{1}d^3k_{2}\sum_{\epsilon_{1},\epsilon_{2}}\left[\hat{\mathbf e}_{\alpha}\cdot\epsilon_{1}(\vec{k}_{1})\right]\left[\hat{\mathbf e}_{\alpha}\cdot\epsilon_{2}(\vec{k}_{2})\right]\nonumber\\
& &\times\left(\kappa^{\star}(\omega_{1})a_{\vec{k}_{1},\epsilon_{1}}^{\dagger}e^{i\omega_{1}t}+\kappa(\omega_{1})a_{\vec{k}_{1},\epsilon_{1}}e^{-i\omega_{1}t}\right)
\left(\kappa^{\star}(\omega_{2})a_{\vec{k}_{2},\epsilon_{2}}^{\dagger}+\kappa(\omega_{2})a_{\vec{k}_{2},\epsilon_{2}}\right)e^{-\beta\sum_{\lambda}n_{\lambda}\omega_{\lambda}}|\{n_{\lambda}\}\ket .
\end{eqnarray}

The expectation value of any operator that does not preserve photon number is zero for each basis state $|\{n_\lambda\}\ket$. Hence, all terms of the type $aa$ and $a^{\dagger}a^{\dagger}$ vanish. The $a^\dagger a$ contribution is
\begin{eqnarray}
\mathcal{J}_{\alpha\alpha}(t)\left\{ a^{\dagger}a\right\}  & = & \frac{1}{Z(\beta)}\int d^3k_{1}d^3k_{2}\sum_{\epsilon_{1},\epsilon_{2}}\left[\hat{\mathbf e}_{\alpha}\cdot\epsilon_{1}(\vec{k}_{1})\right]\left[\hat{\mathbf e}_{\alpha}\cdot\epsilon_{2}(\vec{k}_{2})\right]\kappa^{\star}(\omega_{1})\kappa(\omega_{2})e^{i\omega_{1}t}\nonumber\\
& & \times \sum_{\{n_{\lambda}\}}e^{-\beta\sum_{\lambda}n_{\lambda}\omega_{\lambda}}\bra\{n_{\lambda}\}|a_{\vec{k}_{1},\epsilon_{1}}^{\dagger}a_{\vec{k}_{2},\epsilon_{2}}|\{n_{\lambda}\}\ket\\
 & = & \frac{1}{Z(\beta)}\int d^3k\sum_{\epsilon}\left[\hat{\mathbf e}_{\alpha}\cdot\epsilon(\vec{k})\right]\left[\hat{\mathbf e}_{\alpha}\cdot\epsilon(\vec{k})\right]|\kappa(\omega)|^{2}e^{i\omega t}\nonumber\\
 & & \times \sum_{\{n_{\lambda}\}\backslash n_{\vec{k},\epsilon}}e^{-\beta\sum_{\lambda\backslash(\vec{k},\epsilon)}n_{\lambda}\omega_{\lambda}}\sum_{n_{\vec{k},\epsilon}}e^{-\beta n_{\vec{k},\epsilon}\omega}n_{\vec{k},\epsilon}\\
 & = & \int d^3k\sum_{\epsilon}\left[\hat{\mathbf e}_{\alpha}\cdot\epsilon(\vec{k})\right]\left[\hat{\mathbf e}_{\alpha}\cdot\epsilon(\vec{k})\right]|\kappa(\omega)|^{2}e^{i\omega t}\frac{1}{e^{\beta\omega}-1}.
\end{eqnarray}
The $aa^\dagger$ contribution is similarly
\begin{eqnarray}
\mathcal{J}_{\alpha\alpha}(t)\left\{ aa^{\dagger}\right\}  & = & \frac{1}{Z(\beta)}\int d^3k_{1}d^3k_{2}\sum_{\epsilon_{1},\epsilon_{2}}\left[\hat{\mathbf e}_{\alpha}\cdot\epsilon_{1}(\vec{k}_{1})\right]\left[\hat{\mathbf e}_{\alpha}\cdot\epsilon_{2}(\vec{k}_{2})\right]\kappa(\omega_{1})\kappa^{\star}(\omega_{2})e^{-i\omega_{1}t}\nonumber\\
& &\times \sum_{\{n_{\lambda}\}}e^{-\beta\sum_{\lambda}n_{\lambda}\omega_{\lambda}}\bra\{n_{\lambda}\}|a_{\vec{k}_{1},\epsilon_{1}}a_{\vec{k}_{2},\epsilon_{2}}^{\dagger}|\{n_{\lambda}\}\ket\\
 & = & \frac{1}{Z(\beta)}\int d^3k_{1}d^3k_{2}\sum_{\epsilon_{1},\epsilon_{2}}\left[\hat{\mathbf e}_{\alpha}\cdot\epsilon_{1}(\vec{k}_{1})\right]\left[\hat{\mathbf e}_{\alpha}\cdot\epsilon_{2}(\vec{k}_{2})\right]\kappa(\omega_{1})\kappa^{\star}(\omega_{2})e^{-i\omega_{1}t}\nonumber\\
& &\times \sum_{\{n_{\lambda}\}}e^{-\beta\sum_{\lambda}n_{\lambda}\omega_{\lambda}}\bra\{n_{\lambda}\}|\delta_{\vec{k}_{1},\vec{k}_{2}}\delta_{\epsilon_{1},\epsilon_{2}}+a_{\vec{k}_{2},\epsilon_{2}}^{\dagger}a_{\vec{k}_{1},\epsilon_{1}}|\{n_{\lambda}\}\ket\\
 & = & \int d^3k\sum_{\epsilon}\left[\hat{\mathbf e}_{\alpha}\cdot\epsilon(\vec{k})\right]\left[\hat{\mathbf e}_{\alpha}\cdot\epsilon(\vec{k})\right]|\kappa(\omega)|^{2}e^{-i\omega t}\left(1+\frac{1}{e^{\beta\omega}-1}\right).
\end{eqnarray}

The polarization sums are evaluated using the transverse completeness relation
\begin{equation}
\sum_{\epsilon}
\left(\hat{\mathbf e}_{i}\cdot\epsilon(\vec{k})\right)
\left(\hat{\mathbf e}_{j}\cdot\epsilon(\vec{k})\right)^{\star}
=\delta_{ij}-\hat{k}_{i}\hat{k}_{j}.
\end{equation}
For $i=j$, isotropic angular integration gives the factor $8\pi/3$; for $i\neq j$, the angular integral vanishes. Thus, the diagonal terms of the bath correlator function are
\begin{eqnarray}
\mathcal{J}_{\alpha\alpha}(t) & = & \frac{8\pi}{3c^{3}}\int_{0}^{\infty}d\omega\,|\kappa(\omega)|^{2}\omega^{2}\left[\frac{1}{e^{\beta\omega}-1}e^{i\omega t}+\left(1+\frac{1}{e^{\beta\omega}-1}\right)e^{-i\omega t}\right],\\
\mathcal{J}_{\alpha\alpha}(\omega^\prime) & = & \frac{8\pi}{3c^{3}}\int_{0}^{\infty}d\omega\,|\kappa(\omega)|^{2}\omega^{2}\left[\frac{1}{e^{\beta\omega}-1}\delta(\omega^\prime+\omega)+\left(1+\frac{1}{e^{\beta\omega}-1}\right)\delta(\omega^\prime-\omega)\right]\nonumber\\
\implies \mathcal{J}_{\alpha\alpha}(\omega^\prime)=\mathcal{J}(\omega^\prime) & = & \frac{8\pi}{3c^{3}}(\omega^\prime)^{2}\left[|\kappa(-\omega^\prime)|^{2}\frac{1}{e^{-\beta\omega^\prime}-1}\Theta(-\omega^\prime)+|\kappa(\omega^\prime)|^{2}\left(1+\frac{1}{e^{\beta\omega^\prime}-1}\right)\Theta(\omega^\prime)\right],
\end{eqnarray}
where $\mathcal{J}(\omega^\prime)$ is the scalar bath spectral density and $\Theta$ is the Heaviside step function. The off-diagonal bath correlators are obtained in the same manner:
\begin{eqnarray}
\mathcal{J}_{\alpha\beta}(t) & = & \int d^3k\sum_{\epsilon}\left[\hat{\mathbf e}_{\alpha}\cdot\epsilon(\vec{k})\right]\left[\hat{\mathbf e}_{\beta}\cdot\epsilon(\vec{k})\right]^{\star}|\kappa(\omega)|^{2}\left[e^{i\omega t}\frac{1}{e^{\beta\omega}-1}+e^{-i\omega t}\left(1+\frac{1}{e^{\beta\omega}-1}\right)\right]\nonumber\\
& \propto & \int d\Omega_{\vec{k}}\left(\delta_{\alpha\beta}-\hat{k}_{\alpha}\hat{k}_{\beta}\right)=0,\qquad \alpha\neq\beta.
\end{eqnarray}
The last equality follows from isotropy: the angular average of $\hat{k}_{\alpha}\hat{k}_{\beta}$ is proportional to $\delta_{\alpha\beta}$, so it vanishes when $\alpha\neq\beta$.
Hence the matrix-valued bath correlator is proportional to the identity in the Cartesian noise labels. The jump correlator is therefore also diagonal, with matrix elements $g_{\alpha \alpha}(\omega)=\sqrt{\mathcal{J}_{\alpha \alpha}(\omega)/2\pi}$. This diagonal structure is what allows the three Cartesian noise channels to be rotated independently into spherical channels.

We write $\omega_{mn}=E_n-E_m$, so that $\omega_{mn}>0$ for a transition $|n\ket\to |m\ket$ that lowers the system energy. The unitary rotation from Cartesian components to spherical tensor components is
\begin{equation}
X_{+1}= -\frac{X_x+iX_y}{\sqrt{2}},\qquad
X_{-1}= \frac{X_x-iX_y}{\sqrt{2}},\qquad
X_0=X_z,
\end{equation}
up to convention-dependent phases. With this spherical-basis convention, the jump operators can be written as
\begin{equation}\label{eqn:jumps-spherical-appx}
L_q=
\sqrt{2\pi\gamma}
\sum_{m,n}
\sqrt{\mathcal{J}(\omega_{mn})}
\bra m|X_q|n\ket
|m\ket\bra n|,
\qquad q\in\{-1,0,+1\}.
\end{equation}
Here $\gamma$ is an effective overall decay-rate scale, and $q$ labels the emitted-photon angular-momentum channel, the emitted photon carrying angular momentum $-q$. This is the same channel label denoted by $\delta$ in the main text.

In the zero-temperature limit, the thermal state reduces to the vacuum state $|\mathrm{vac}\ket\bra\mathrm{vac}|$, causing the $\mathcal{J}_{\alpha\alpha}(t)\{ a^{\dagger}a\}$ term to vanish. As a result, only the positive-frequency terms in $\mathcal{J}(\omega)$ survive, so transitions occur only from higher-energy states to lower-energy states. The jump operators therefore have the same form as in~\cref{eqn:jumps-spherical-appx} with
\begin{equation}
\mathcal{J}(\omega)=\frac{8\pi}{3c^{3}}\omega^{2}|\kappa(\omega)|^{2}\Theta(\omega).
\end{equation}
With $\mathcal{E}_{\omega}\propto\sqrt{\omega}$ this is the familiar free-space result $\mathcal{J}(\omega)\propto\omega^{3}$. The mode sum has been converted to $\int d^3k$ with the density-of-states factor $V/(2\pi)^3$ absorbed into $\kappa$.
For angular-momentum eigenstates, the Wigner--Eckart theorem gives
\begin{equation}
\bra J^\prime,n|X_q|J,m\ket
\propto
C^{J^\prime,n}_{1,q;J,m},
\end{equation}
up to convention-dependent normalization and phase choices. For the effective model used in the main text, we restrict to the downward angular-momentum branch $J^\prime=J-1$ (the branch $J^\prime=J$ has $\omega_{mn}=0$, where $\mathcal{J}(0)=0$, while $J^\prime=J+1$ raises the energy and is absent at zero temperature) and absorb species-dependent reduced matrix elements, radial factors, field-normalization constants, and spectral-density factors into the effective parameters $\gamma$ and $f(J)$. This recovers the simplified jump operators quoted in the main text,
\begin{equation}
L_{\delta}
=
\sqrt{\gamma}
\sum_{J,|m|\leq J}
f(J)
C^{J-1,m+\delta}_{1,\delta;J,m}
|J-1,m+\delta\ket\bra J,m|,
\qquad \delta\in\{-1,0,+1\}.
\end{equation}
Here $C$ denotes the Clebsch--Gordan coefficient enforcing the angular-momentum selection rules, and $f(J)$ absorbs reduced matrix elements, radial factors, and the relevant spectral-density normalization. The degeneracy-breaking term $H_\Delta$ introduced in the main text is not part of the spontaneous-decay Hamiltonian itself; it is a phenomenological Hamiltonian used to study how lifted degeneracies make distinct decay paths more distinguishable.

%% ======================================================================
%% S4: Fast Checking in Atomic and Molecular Systems
%% ======================================================================
\appsection{Fast Checking in Atomic and Molecular Systems}\label{sm:fast}
This section justifies the fast-checking claim of the main text for the concrete angular-momentum model used there. As in the four-level example, the relevant question is what happens during one short check-and-recovery interval. Each check measures only whether the state has left the code manifold $J_0$, that is, whether a decay to $J_0-1$ has occurred. Null outcomes produce deterministic phases that can be tracked if the lifting is calibrated, while a detected single decay is decoded after at most one checking interval of additional evolution. Events with two or more decays in the same interval have probability $O((\gamma\tau)^2)$ per interval, and therefore vanish over a fixed total time as $\tau\rightarrow0$. Thus, on any finite-dimensional code support with bounded lifting splittings, the residual distinguishability associated with a detected correctable decay vanishes in the continuous-checking limit. The system is described by the following Hamiltonian and jump operators:
\begin{eqnarray*}
H & = & \sum_{J,m}\left[\omega_{0}J(J+1)+\Delta(J,m)\right]|J,m\ket\bra J,m|,\\
L_{1} & = & \sqrt{\gamma}\sum_{{J\geq1},|m|\leq J}\sqrt{\frac{2J+1}{2J-1}}\,C_{1,1;J,m}^{J-1,m+1}|J-1,m+1\ket\bra J,m|,\\
L_{-1} & = & \sqrt{\gamma}\sum_{{J\geq1},|m|\leq J}\sqrt{\frac{2J+1}{2J-1}}\,C_{1,-1;J,m}^{J-1,m-1}|J-1,m-1\ket\bra J,m|,\\
L_{0} & = & \sqrt{\gamma}\sum_{{J\geq1},|m|\leq J}\sqrt{\frac{2J+1}{2J-1}}\,C_{1,0;J,m}^{J-1,m}|J-1,m\ket\bra J,m|.
\end{eqnarray*}

Here the level scaling $\eta_J$ of \cref{eqn:atomic-hamiltonian_appx} is specialized to $\eta_J=J(J+1)$. The form of the lift here is more general than the main text, which is recovered by setting $\Delta(J,m)=\Delta_{0}Jm$, with $\Delta_0$ the lifting strength $\Delta$ of \cref{eqn:atomic-non-degeneracy}. For the short-time expansion below, assume that the measurement time is shorter than both the error timescale and the lifting timescale, i.e., $\tau \ll 1/\gamma$ and $\tau\ll 1/\max|\Delta(J,m)|$, the maximum taken over the occupied states (for $\Delta(J,m)=\Delta_0Jm$ and a code hosted in $J_0$ this is $\tau\ll1/\Delta_0J_0^2$). No condition on $\omega_{0}$ is needed since $\omega_{0}J(J+1)$ is a global phase within each $J$ manifold. Then the approximate Kraus operators of the system~\cite{albert2018lindbladians} are
\begin{eqnarray*}
K_{+1} & = & \sqrt{\gamma \tau}\sum_{{J\geq1},|m|\leq J}\sqrt{\frac{2J+1}{2J-1}}\,C_{1,1;J,m}^{J-1,m+1}|J-1,m+1\ket\bra J,m|,\\
K_{-1} & = & \sqrt{\gamma \tau}\sum_{{J\geq1},|m|\leq J}\sqrt{\frac{2J+1}{2J-1}}\,C_{1,-1;J,m}^{J-1,m-1}|J-1,m-1\ket\bra J,m|,\\
K_{0} & = & \sqrt{\gamma \tau}\sum_{{J\geq1},|m|\leq J}\sqrt{\frac{2J+1}{2J-1}}\,C_{1,0;J,m}^{J-1,m}|J-1,m\ket\bra J,m|,\\
K_{\mathrm{null}} & = & 1+(-iH+V)\tau,
\end{eqnarray*}
where
\begin{eqnarray*}
V & = & -\frac{1}{2}\left(L_{1}^{\dagger}L_{1}+L_{-1}^{\dagger}L_{-1}+L_{0}^{\dagger}L_{0}\right).
\end{eqnarray*}
Using the Clebsch--Gordan identity below,
\begin{eqnarray}
L_{\delta}^{\dagger}L_{\delta} & = & \gamma\sum_{{J\geq1},|m|\leq J}\frac{2J+1}{2J-1}\left|C_{1,\delta;J,m}^{J-1,m+\delta}\right|^{2}|J,m\ket\bra J,m|\nonumber\\
\implies V & = & -\frac{\gamma}{2}\sum_{J\geq 1,m}|J,m\ket\bra J,m|,
\end{eqnarray}
where we used the identity
\begin{eqnarray}
\sum_{\delta\in\{0,\pm 1\}}\left|C_{1,\delta;J,m}^{J-1,m+\delta}\right|^{2} & = & \frac{2J-1}{2J+1}.
\end{eqnarray}
The identity, and hence all sums above, hold for $J\geq1$. The $J=0$ ground manifold has no downward transition and is excluded throughout.
Consequently, $K_{\mathrm{null}}$ has the simple form
\begin{eqnarray}
K_{\mathrm{null}} & = & 1-i\sum_{J,m}\left[\omega_{0}J(J+1)+\Delta(J,m)\right]|J,m\ket\bra J,m|\,\tau-\frac{\gamma}{2}\sum_{J\geq 1,m}|J,m\ket\bra J,m|\,\tau.
\end{eqnarray}

For a fast checking time $\tau\ll 1/\gamma,\,1/\max|\Delta(J,m)|$, the state evolves as
\begin{eqnarray}
\rho_{0} & \rightarrow & K_{\mathrm{null}}\rho_{0}K_{\mathrm{null}}^{\dagger}+\sum_{\delta\in\{0,\pm1\}}K_{\delta}\rho_{0}K_{\delta}^{\dagger}.
\end{eqnarray}
Now assume that the initial state lies within a single $J$ manifold, say $J=J_{0}$ with $J_{0}\geq1$ (a $J_{0}=0$ state supports no decay and evolves trivially), so that $\Pi_{J_{0}}\rho_{0}\Pi_{J_{0}}=\rho_{0}$. We define a projective syndrome measurement via the operators $M_{\mathrm{null}} = \Pi_{J_0}$ and $M_{\mathrm{decay}} = I-M_{\mathrm{null}}$. The state after a fast measurement with null syndrome is
\begin{equation}
\dfrac{\Pi_{J_{0}}K_{\mathrm{null}}\rho_{0}K_{\mathrm{null}}^{\dagger}\Pi_{J_{0}}}{\Tr\left[\Pi_{J_{0}}K_{\mathrm{null}}\rho_{0}K_{\mathrm{null}}^{\dagger}\right]}=\dfrac{K_{\mathrm{null}}\rho_{0}K_{\mathrm{null}}^{\dagger}}{\Tr\left[K_{\mathrm{null}}\rho_{0}K_{\mathrm{null}}^{\dagger}\right]}.
\end{equation}
The $K_{\delta}$ terms with $\delta\in\{0,\pm1\}$ are absent because they necessarily
change $J$ and therefore vanish after projection
onto the $J_{0}$ manifold. Since $K_{\mathrm{null}}$ is a diagonal operator,
one obtains
\begin{eqnarray*}
K_{\mathrm{null}}\rho_{0}K_{\mathrm{null}}^\dagger & = & \rho_{0}+\tau(-iH+V)\rho_{0}+\tau\rho_{0}(iH+V)\\
 & = & \rho_{0}+i[\rho_{0},H]\tau+V\rho_{0}\tau+\rho_{0}V\tau.
\end{eqnarray*}

In the perfectly degenerate case with $\Delta(J,m)=0$, $K_{\mathrm{null}}$ has no $m$ dependence and hence $[H,\rho_{0}]=0$. Thus, for $\rho_{0}=\sum_{m,m^{\prime}}\rho_{m,m^{\prime}}|J_{0},m\ket\bra J_{0},m^{\prime}|$,
\begin{eqnarray}
K_{\mathrm{null}}\rho_{0}K_{\mathrm{null}}^\dagger & = & \left(1-\gamma\tau\right)\rho_{0}+O(\tau^{2})\nonumber\\
\implies\dfrac{K_{\mathrm{null}}\rho_{0}K_{\mathrm{null}}^{\dagger}}{\Tr\left[K_{\mathrm{null}}\rho_{0}K_{\mathrm{null}}^{\dagger}\right]} & = & \rho_{0}.
\end{eqnarray}

As long as the state does not decay, the measurement alone is enough to keep the code state intact in the degenerate case, and the clock can effectively be reset to $t=0$. When the code state does decay, the resulting state in the $J_{0}-1$ manifold can be written as
\begin{eqnarray*}
\rho(\tau) & = & \dfrac{\sum_{\delta\in\{0,\pm1\}}K_{\delta}\rho_{0}K_{\delta}^{\dagger}}{\Tr\left[\sum_{\delta\in\{0,\pm1\}}K_{\delta}\rho_{0}K_{\delta}^{\dagger}\right]}.
\end{eqnarray*}

Abbreviating the matrix element $\bra J,m|\rho|J,m^{\prime}\ket$
as $\rho_{m,m^{\prime}}^{(J)}$, the unnormalized decayed state is
\begin{eqnarray}
\sum_{\delta\in\{0,\pm1\}}K_\delta\rho_0K_\delta^\dagger
& = & \frac{(2J_0+1)\,\gamma\tau}{2J_0-1}
\sum_{m,m^{\prime}}\sum_{\delta\in\{0,\pm1\}}
C_{1,\delta;J_0,m}^{J_0-1,m+\delta}
C_{1,\delta;J_0,m^{\prime}}^{\star J_0-1,m^{\prime}+\delta}
\rho_{m,m^{\prime}}^{(J_0)}(0)
\nonumber\\
& & \times
|J_0-1,m+\delta\ket\bra J_0-1,m^{\prime}+\delta|,\label{eqn:decayed-unnormalized}\\
\Tr\left[\sum_{\delta\in\{0,\pm1\}}K_{\delta}\rho_{0}K_{\delta}^{\dagger}\right] & = & \gamma \tau\,\frac{2J_{0}+1}{2J_{0}-1}\sum_m\rho_{m,m}^{(J_0)}(0)\sum_{\delta\in\{0,\pm 1\}}\left|C_{1,\delta;J_{0},m}^{J_{0}-1,m+\delta}\right|^{2} \nonumber \\
 & = & \gamma \tau\label{eqn:decayed-trace}\\
\implies\rho(\tau) & = & \dfrac{\sum_{\delta\in\{0,\pm1\}}K_{\delta}\rho_{0}K_{\delta}^{\dagger}}{\Tr\left[\sum_{\delta\in\{0,\pm1\}}K_{\delta}\rho_{0}K_{\delta}^{\dagger}\right]}\nonumber\\
 & = & \frac{2J_{0}+1}{2J_{0}-1}\sum_{m,m^{\prime}}\sum_{\delta\in\{0,\pm 1\}}C_{1,\delta;J_{0},m}^{J_{0}-1,m+\delta}C_{1,\delta;J_{0},m^{\prime}}^{\star J_{0}-1,m^{\prime}+\delta}\nonumber \\
 & & \times\rho_{m,m^{\prime}}^{(J_0)}(0)|J_{0}-1,m+\delta\ket\bra J_{0}-1,m^{\prime}+\delta|.\label{eqn:decayed-normalized}
\end{eqnarray}

This has the form of a perfectly recoverable error state when $\rho_{0}$ is an encoded state of a code capable of correcting such transitions, such as the \AE\ or Aydin--Barg codes.

In the case of nonzero lifting strength, i.e.~$\Delta(J,m)=Jm\Delta_{0}$ with $\Delta_{0}\neq0$, $K_{\mathrm{null}}$ acquires an $m$ dependence. Within the $J_0$ manifold, $i[\rho_0,H]_{m,m^\prime}=-i\Delta_{0}J_{0}\left(m-m^{\prime}\right)\rho_{m,m^\prime}^{(J_0)}$
and
\begin{eqnarray}
\bra J_{0},m|K_{\mathrm{null}}\rho_{0}K_{\mathrm{null}}^{\dagger}|J_{0},m^{\prime}\ket & = & \left[1+\left(-i\Delta_{0}J_{0}(m-m^{\prime})-\gamma\right)\tau\right]\rho_{m,m^{\prime}}^{(J_{0})}(0)+O(\tau^{2}),\\
\Tr\left[K_{\mathrm{null}}\rho_{0}K_{\mathrm{null}}^{\dagger}\right] & = & \left(1-\gamma\tau\right)+O(\tau^{2})\\
\implies\rho_{m,m^{\prime}}^{(J_{0})}(\tau) & = & \rho_{m,m^{\prime}}^{(J_{0})}(0)\left(1-\frac{i\Delta_{0}J_{0}(m-m^{\prime})\tau}{1-\gamma\tau}\right) \nonumber\\
& \sim & \rho_{m,m^{\prime}}^{(J_{0})}(0)\left(1-i\Delta_{0}J_{0}(m-m^{\prime})\tau\right)+O(\tau^{2}).
\end{eqnarray}

The null branch thus undergoes the rotation $e^{-i\theta J_z}$, $\theta=\Delta_0J_0\tau$, in each interval. Iterating it gives $\rho_{m,m^{\prime}}(t)=\rho_{m,m^{\prime}}(0)e^{-i\Delta_{0}J_{0}(m-m^{\prime})t}$ at time $t$, which coincides with the uninterrupted no-jump evolution: since $\Pi_{J_{0}}$ commutes with $K_{\mathrm{null}}$, the projections post-select the null branch without modifying the decay rate, which remains $\gamma$. These phases are deterministic and are tracked and compensated from the elapsed evolution time, which requires $\Delta(J,m)$ to be known. 

Thus, once these phases are compensated, the non-decayed state is returned to the code state at every check.
The decayed branch after a check at time $t$ is again given by \crefrange{eqn:decayed-unnormalized}{eqn:decayed-normalized} with $\rho^{(J_{0})}_{m,m^{\prime}}(0)\to\rho_{m,m^{\prime}}(t)$.

\paragraph{Remark: uncalibrated lifts.} A possible way to avoid calibration is to augment each check so that it also resolves the $J_z$ syndrome, distinguishing $\mathcal{C}$ from $J_z\mathcal{C}$ within the $J_0$ manifold (these are orthogonal since $\bra J_z\ket=0$ for both codes), and on the $J_z\mathcal{C}$ outcome applies the isometry $J_z|\bar\mu\ket/\sqrt{\bra J_z^2\ket}\mapsto|\bar\mu\ket$. This isometry is well defined because the Knill--Laflamme conditions give $\bra\bar\mu|J_z^2|\bar\nu\ket=\bra J_z^2\ket\delta_{\mu\nu}$. Since $J_z$ is a rank-$1$ tensor operator and the \AE\ and Aydin--Barg codes satisfy these conditions for $\{1,J_z,J_\pm\}$~\cite{jain2024absorptionemission,aydin2025class}, such an augmented check corrects the term linear in the rotation angle $\theta$. Establishing the accumulated error from the remaining higher-order terms requires an analysis of the complete repeated-check channel, so we leave the performance of this uncalibrated extension for future work.

Rapid checking effectively erases the $J$ dependence of the degeneracy lift after a decay, because the system has at most one checking interval to accumulate phase while evolving in the $J-1$ manifold before the decay is detected. If the functional form of $\Delta(J,m)$ is known, the deterministic phases in both the null and detected-decay branches can be incorporated into the conditional recovery. In the detected-decay branch, the phase also depends on the unobserved jump time within the checking interval. Calibration can therefore remove its coherent part, but not the residual dephasing obtained after averaging over possible jump times. This residual vanishes with $\tau$ because the jump time is confined to an increasingly short interval. The augmented $J_z$-syndrome check described above may offer an alternative when the lift is uncalibrated, but its higher-order residual errors under repeated checking remain to be analyzed. With calibrated phase compensation, the single-decay branch has the same correctable form as in the degenerate case, up to residual terms that vanish with $\tau$. The important point is that the distinguishability accumulated between the decay and its detection disappears in the continuous-checking limit. Multi-decay events within one interval are negligible in the same limit, as noted above. Since the number of detected decays over a fixed total time is set by the physical decay rate rather than by the checking frequency, repeated check-and-recovery cycles approach the ideal error-correction picture as $\tau\rightarrow0$.

\begin{figure}[h!tbp]
    \centering
    \includegraphics[width = 0.5\columnwidth]{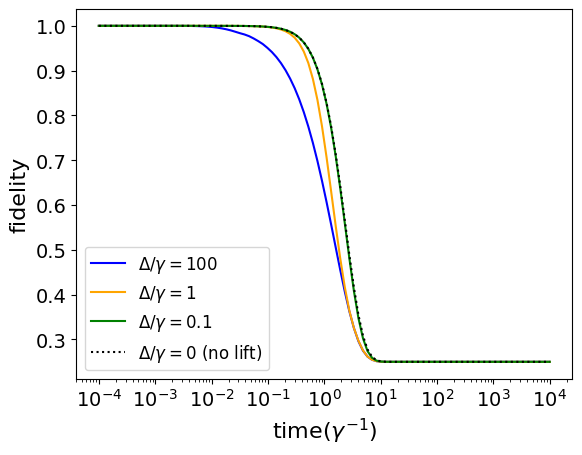}
    \caption{Evolution of the transpose fidelity $F_{\mathrm{tr}}$ of the Aydin--Barg code, hosted in $J=4$, at lifting strengths $\Delta/\gamma=0$, $0.1$, $1$, and $100$. The long-time plateau at $F_{\mathrm{tr}}=1/4$ corresponds to complete loss of the logical information. This code, from Ref.~\cite{aydin2025class}, corrects any rank-$1$ transition error, $J\to J^\prime$ with $|J-J^\prime|\le1$ and $m\to m^\prime$ with $|m-m^\prime|\le1$. At zero temperature only the $J\to J-1$ transitions of \cref{eqn:jumps-atom-main-text} occur.}
   \label{fig:arda-performance-diff-delta}
\end{figure}

%% ======================================================================
%% S5: Additional Simulations
%% ======================================================================
\appsection{Additional Simulations}\label{sm:sims}
In this section, we collect additional numerical checks that support the trends emphasized in the main text. As a first check, we compute the evolution of code performance with time for different strengths of the degeneracy lift. We show this in~\cref{fig:arda-performance-diff-delta}, where we plot the fidelity versus time for the Aydin--Barg code, which corrects up to rank-$1$ errors. The fidelity decays more rapidly for larger lifting strengths, confirming that making the states, and hence the error processes, more distinguishable is detrimental to QEC performance.

\begin{figure}[h!tbp]
    \centering
    \begin{subfigure}[t]{0.45\textwidth}
        \caption{}
        \begin{tabular}{cc}
\hline 
\hline
 code & curve\tabularnewline
 
\hline 
minimal & $\quad\quad \tau(\gamma)\approx{1.3\times 10^{-4}}/{\gamma}\quad\quad$ \tabularnewline
detection & $\tau(\gamma)\approx{2.0\times 10^{-4}}/{\gamma}$
\tabularnewline
Aydin--Barg & $\tau(\gamma)\approx{0.020}/{\gamma}$ 
\tabularnewline
\AE & $\tau(\gamma)\approx{0.026}/{\gamma}$ \tabularnewline
\hline 
\hline
\end{tabular}
    \end{subfigure}
    \hfill
    \begin{subfigure}[t]{0.45\textwidth}
        \caption{}
        \centering
        \includegraphics[width =\columnwidth]{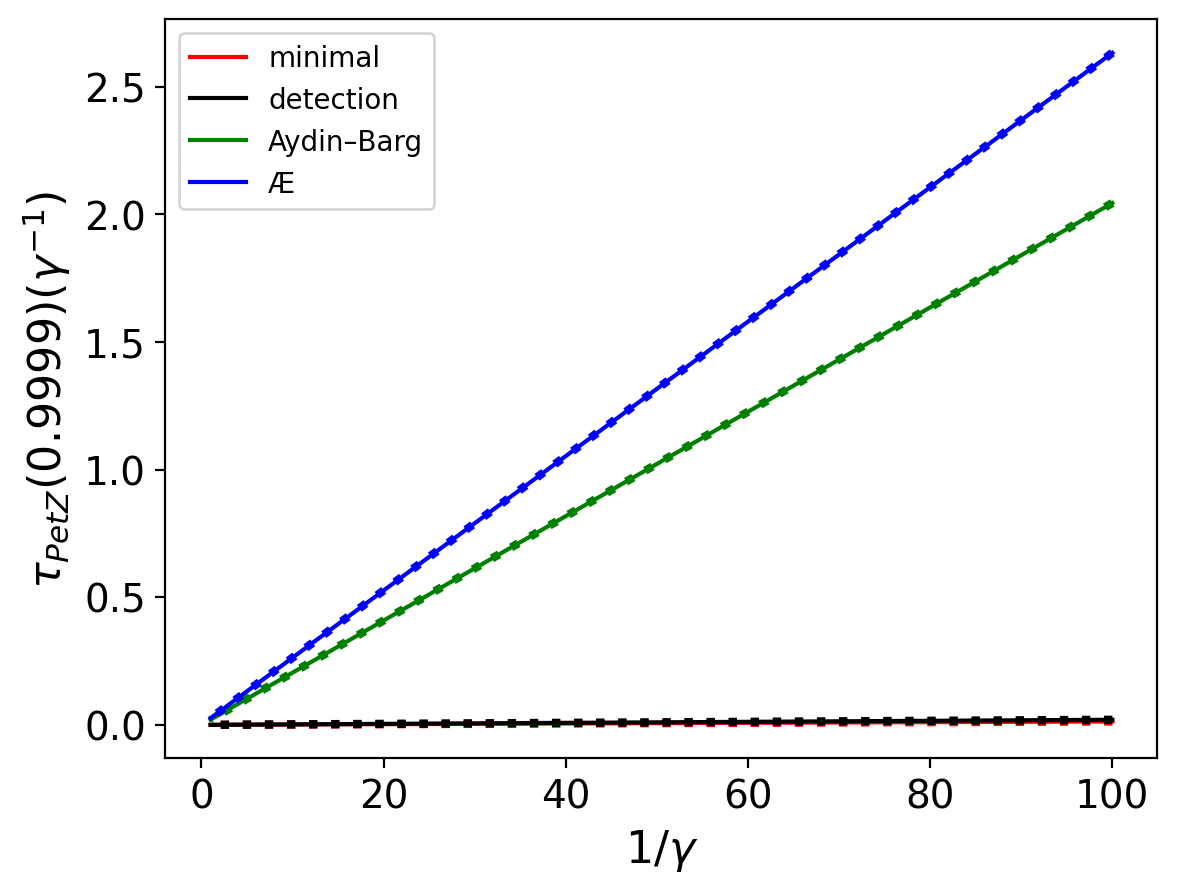}
    \end{subfigure}
    
    \caption{Checking time as a function of the error rate in the absence of degeneracy lifting (solid curves), at target fidelity $F_0=0.9999$. The checking time is inversely proportional to the error rate $\gamma$, as shown by the linear fits in the table and plotted as dotted curves. The two baselines have closed forms: the minimal code obeys $1-F_{\mathrm{tr}}=\tfrac34 p_{\mathrm{decay}}$ exactly, so $\tau=-\gamma^{-1}\ln[1-\tfrac43(1-F_0)]\simeq4(1-F_0)/(3\gamma)$, and the detection code obeys $1-F_{\mathrm{tr}}\simeq\gamma t/2$, so $\tau\simeq2(1-F_0)/\gamma$. The Aydin--Barg and \AE\ values are numerical fits.}
    \label{fig:checking_time_gamma_varying}
\end{figure}

It is also useful to examine how the checking time depends on the error rate itself in the absence of distinguishability effects. We study this in~\cref{fig:checking_time_gamma_varying}, where we plot the checking time as a function of the error rate $\gamma$ for several encodings. In the absence of degeneracy lifting, the checking time is inversely proportional to the error rate, so the syndrome-checking interval must shrink proportionally as the physical error rate grows if one wishes to maintain the same performance. Among the codes considered here, the \AE\ code achieves the largest checking time.

\paragraph{Numerical method.} The channel has a weak symmetry generated by the rotation $R_{\theta}=e^{-i\theta J_{z}}$: since $H_{\mathcal{S}}$ and $H_\Delta$ are diagonal in the $|J,m\ket$ basis and $R_{\theta}L_{\delta}R_{\theta}^{\dagger}=e^{-i\theta\delta}L_{\delta}$, the Lindbladian is invariant under conjugation by $R_\theta$~\cite{albert2014symmetries}. It is therefore block diagonal in the sectors $\mathcal{H}_r=\{|J,m\ket\bra J^\prime,m^\prime|\,;\,m-m^\prime=r\}$, which are mutually decoupled. This decomposition can be used to evolve the sectors separately and in parallel. The systems considered here are small enough that we instead exponentiate the full Lindbladian directly and extract the Kraus operators from the Choi matrix of the resulting channel~\cite{watrous2018theory}.

\onecolumngrid %% If need to put in tables or single column content

\newpage

\twocolumngrid
\let\o\SJtextoslash % bib entries such as M{\o}lmer need the text "ø", not \omega
\bibliography{refs.bib}

\end{document}